%% file: report.tex
\documentclass[11pt,a4paper]{llncs}

\usepackage[usenames,dvipsnames]{color}
\usepackage[francais,english]{babel}
\usepackage[T1]{fontenc}
\usepackage[utf8]{inputenc}
\usepackage{pslatex}
\usepackage{amsmath}
\usepackage{epsfig}
\usepackage{wrapfig}
\usepackage{paralist}
\usepackage{graphics}
\usepackage{stmaryrd}
\usepackage{txfonts}
\usepackage{framed}
\usepackage{makecell}
\usepackage{url}
\usepackage[colorlinks]{hyperref}
\usepackage{proof}

\usepackage{algorithm}
\usepackage{algorithmicx}
\usepackage[noend]{algpseudocode}

\usepackage[final]{commenting}

\newif\ifLongVersion\LongVersiontrue

\include{commands}

\declareauthor{ri}{Radu}{blue}
\declareauthor{np}{Nicolas}{red}
\declareauthor{me}{Mnacho}{green!70!black}

\begin{document}
%%%%%%%%%%%%%%%%%%%%%%%%%%%%%%%%%%%%%%%%%%%%%%%%%%%%%%%%%%%%%%%%%%%%%%%%%%%%%%%

\title{The Complexity of Prenex Separation Logic with One Selector}
\author{M. Echenim\inst{1}, R. Iosif\inst{2} and N. Peltier\inst{1}}
 \institute{Univ. Grenoble Alpes, CNRS, LIG, F-38000 Grenoble France
   \and Univ. Grenoble Alpes, CNRS, VERIMAG, F-38000 Grenoble France}

\maketitle 

\input abstract

\input body

\bibliographystyle{abbrv}
\bibliography{refs}

%%%%%%%%%%%%%%%%%%%%%%%%%%%%%%%%%%%%%%%%%%%%%%%%%%%%%%%%%%%%%%%%%%%%%%%%%%%%%%%
\end{document}

%% file: commands.tex
\newcommand{\bigO}{\mathcal{O}}

\newcommand{\pspace}{\textsf{PSPACE}}
\newcommand{\ptime}{\textsf{P}}

\newcommand{\bigospace}[1]{\mathsf{SPACE}({#1})}

\newcommand{\set}[1]{\left\{ #1 \right\}}
\newcommand{\tuple}[1]{\left\langle #1 \right\rangle}
\renewcommand{\vec}[1]{\mathbf #1}

\newcommand\equivalent[2]{$(#1,#2)$-similar}

\newcommand{\defequal}{\stackrel{\scriptscriptstyle{\mathsf{def}}}{=}}

\newcommand{\len}[1]{{|{#1}|}}
\newcommand{\card}[1]{{||{#1}||}}

\newcommand{\nat}{{\bf \mathbb{N}}}
\newcommand{\zed}{{\bf \mathbb{Z}}}

\newcommand{\hoare}[3]{\{#1\}\ {\bf #2}\ \{#3\}}

\renewcommand{\paragraph}[1]{\noindent{\bf #1}}

\newcommand{\teq}{\approx}

\newcommand{\I}{\mathcal{I}}
\newcommand{\J}{\mathcal{J}}

\newcommand{\emp}{\mathsf{emp}}
\newcommand{\wand}{
 \mathrel{\mbox{$\hspace*{-0.03em}\mathord{-}\hspace*{-0.66em}
 \mathord{-}\hspace*{-0.36em}\mathord{*}$\hspace*{-0.005em}}}} % {\multimap}
\newcommand{\seplog}{\mathsf{SL}}

\newcommand{\seplogk}[1]{\seplog^{\!\scriptstyle{#1}}}

\newcommand{\SLtoFO}[1]{\Theta(#1)}
\newcommand{\domh}{\mathfrak{d}}
\newcommand{\funch}{\mathfrak{f}}
\newcommand{\nodesPi}{\mu}
\newcommand{\nodesCh}{\nu}
\newcommand{\rootnode}{{\mathfrak{r}}}
\newcommand{\semf}{\mathfrak{f}}
\newcommand{\propr}{{\cal P}}
\newcommand{\successor}{\mathit{succ}}
\newcommand{\tower}{\mathrm{T}}

\newcommand{\fol}{\mathsf{FO}}
\newcommand{\bsr}{\mathsf{BSR}}
\newcommand{\prenex}{\mathsf{PRE}}

\newcommand{\fv}[1]{\mathsf{var}({#1})}

\newcommand{\dom}{\mathrm{dom}}
\newcommand\Qu{\mathrm{Q}}
\newcommand{\var}[1]{\mathit{fv}(#1)}

\newcommand{\interv}[2]{\llbracket#1\mathrel{{.}\,{.}}\nobreak#2\rrbracket}
\newcommand{\equivs}[2]{\sim_{#1}^{#2}}
\newcommand{\elemh}[1]{\mathrm{elems}(#1)}

\newcommand{\alldistinct}{\mathsf{distinct}}
\newcommand\form[1]{\lambda_{#1}}

\newcommand{\equivfin}{\equiv^{\fincard}}

\newcommand{\img}{\mathrm{img}}

\newcommand{\size}[1]{\mathsf{size}(#1)}
\newcommand{\vars}{\mathsf{Var}}

\renewcommand{\int}{\mathsf{\scriptscriptstyle{i}}}

\newcommand{\maxn}[1]{\mathcal{N}({#1})}

\newcommand{\U}{\mathfrak{U}}
\newcommand{\afunc}{\mathfrak{i}}
\newcommand{\astruct}{\mathcal{S}}
\newcommand{\astore}{\mathfrak{s}}
\newcommand{\aheap}{\mathfrak{h}}
\renewcommand{\iff}{\Leftrightarrow}
\newcommand{\pto}{\hookrightarrow}
\newcommand{\alloc}{\mathsf{alloc}}

\newcommand{\septraction}{\multimap}

\newcommand{\finmt}[1]{\mu^{\scriptscriptstyle{\fincard}}({#1})}

\newcommand{\fincard}{\mathit{fin}}
\newcommand{\infcard}{\mathit{inf}}

\newcommand{\minheap}{\mathrm{hmin}}
\newcommand{\maxheap}{\mathrm{hmax}}

\newcommand{\bounded}{$\mathcal{M}$-bounded}

\newcommand{\Bool}{\mathsf{Bool}}

\newcommand{\isdef}{\defequal}

%% file: abstract.tex
\begin{abstract}
We first show that infinite satisfiability can be reduced to finite
satisfiability for all prenex formulas of Separation Logic with
$k\geq1$ selector fields ($\seplogk{k}$). Second, we show that this
entails the decidability of the finite and infinite satisfiability
problem for the class of prenex formulas of $\seplogk{1}$, by
reduction to the first-order theory of one unary function symbol and
unary predicate symbols. We also prove that the complexity is not
elementary, by reduction from the first-order theory of one unary
function symbol. Finally, we prove that the
Bernays-Sch\"onfinkel-Ramsey fragment of prenex $\seplogk{1}$ formulae
with quantifier prefix in the language $\exists^*\forall^*$
is \pspace-complete. The definition of a complete (hierarchical)
classification of the complexity of prenex $\seplogk{1}$, according to
the quantifier alternation depth is left as an open problem.
\end{abstract}

%% file: body.tex
\section{Introduction}

Separation Logic \cite{IshtiaqOHearn01,Reynolds02} ($\seplog$) is a
logical framework used in program verification to describe properties
of the heap memory, such as the placement of pointer variables within
the topology of complex data structures, such as lists or trees. The
features that make $\seplog$ attractive for program verification are
the ability of defining \begin{inparaenum}[(i)] \item weakest pre- and
  post-condition calculi that capture the semantics of programs with
  pointers, and
\item compositional verification methods, based on the principle of
  local reasoning, which consists of infering separate specifications
  of different parts of a program and combining these specifications
  a~posteriori, in a global verification condition. \end{inparaenum}

The search for automated push-button program verification methods
motivates the need for understanding the decidability, complexity and
expressive power of various dialects thereof, that are used as
assertion languages in Hoare-style proofs \cite{IshtiaqOHearn01}, or
logic-based abstract domains in static analysis \cite{Infer}.

Essentially, one can view $\seplog$ as the first order theory of
the heap using quantification over heap locations, to which two
non-classical connectives are added:~\begin{inparaenum}[(i)]
\item the \emph{separating conjunction} $\phi_1 * \phi_2$, that
  asserts a split of the heap into disjoint heaps satisfying $\phi_1$
  and $\phi_2$ respectively, and
\item the \emph{separating implication} or
\emph{magic wand} $\phi_1 \wand \phi_2$, stating that each extension
of the heap by a heap satisfying $\phi_1$ must satisfy
$\phi_2$.  
\end{inparaenum}

Let us consider the following Hoare triple defining the weakest
precondition of a selector update in a program handling lists, such as
the classical in-place list reversal example \cite{Reynolds02}:
\[\hoare{\exists x ~.~ \mathsf{i} \mapsto x * (\mathsf{i} \mapsto 
\mathsf{j} \wand \phi)}{\mathsf{i.next} = \mathsf{j}}{\phi}\]
A typical verification condition asks whether the weakest precondition
formula is entailed by another precondition $\psi$, generated by a
program verifier or supplied by the user. The entailment $\psi
\rightarrow \exists x ~.~ \mathsf{i} \mapsto x * (\mathsf{i} \mapsto
\mathsf{j} \wand \phi)$ is valid if and only if the formula $\theta = \psi
\wedge \forall x ~.~ \neg (\mathsf{i} \mapsto x * (\mathsf{i} \mapsto
\mathsf{j} \wand \phi))$ is unsatisfiable. 

Assume now that $\phi$ and $\psi$ are formulae of the form $Q_1x_1
\ldots Q_nx_n ~.~ \varphi$, where $Q_1, \ldots, Q_n$ are the first
order quantifiers $\exists$ and $\forall$ and $\varphi$ is
quantifier-free. These formulae are said to be in \emph{prenex}
form. Because the assertions $\mathsf{i} \mapsto x$ and $\mathsf{i}
\mapsto \mathsf{j}$ define \emph{precise} parts of the heap, the
quantifiers of $\phi$ can be hoisted and the entire formula $\theta$
can be written in prenex form, following the result of \cite[Lemma
  3]{OHearnYangReynolds04}.

Deciding the satisfiability of prenex $\seplog$ formulae is thus an
important ingredient for push-button program verification. In general,
unlike first order logic, $\seplog$ formulae do not have a prenex form
because e.g. $\phi * \forall x ~.~ \psi(x) \not\equiv \forall x ~.~
\phi * \psi(x)$ and $\phi \wand \exists x ~.~ \psi(x) \not\equiv
\exists x ~.~ \phi \wand \psi(x)$. Moreover, it was proved that, for
heaps with only one selector, $\seplog$ is undecidable in the presence
of $*$ and $\wand$ (in fact $\seplogk{1}$ is as expressive as second
order logic), whereas the fragment of $\seplog$ without $\wand$ is
decidable but not elementary recursive \cite{BrocheninDemriLozes11}.

In this paper we answer several open problems, by showing that: 
\begin{compactenum}
\item the prenex fragment of $\seplogk{1}$ with $*$ and $\wand$ is
  decidable but not elementary recursive, and
\item the Bernays-Sch\"onfinkel-Ramsey fragment of $\seplogk{1}$ with
  $*$ and $\wand$ is \pspace-complete.
\end{compactenum}

All results in this paper have been obtained using reductions to and
from first order logic with one monadic function symbol, denoted as
$[\mathit{all}, (\omega), (1)]_=$ in
\cite{BorgerGradelGurevich97}. The decidability of this fragment is a
consequence of the celebrated Rabin Tree Theorem \cite{Rabin69}, which
established the decidability of monadic second order logic of the
infinite binary tree (S2S). Furthermore, the $[\mathit{all}, (\omega),
  (1)]_=$ fragment is shown to be nonelementary, by a direct reduction
from domino problems of size equal to a tower of exponentials and,
finally, the $[\exists^*\forall^*, (\omega), (1)]_=$ fragment is
proved to be $\Sigma_2^P$-complete \cite{BorgerGradelGurevich97}.

Essential to our reductions to and from $[\mathit{all}, (\omega),
  (1)]_=$ is a result stating that each quantifier-free $\seplogk{k}$
formula, for $k\geq1$, is equivalent to a boolean combination of
patterns, called \emph{test formulae} \cite{EIP18}. Similar
translations exist for quantifier-free $\seplogk{1}$
\cite{PhD-lozes,BrocheninDemriLozes11} and for $\seplogk{1}$ with one
quantified variable \cite{DemriGalmicheWendlingMery14}. In our
previous work \cite{EIP18}, we have considered both the finite and
infinite satisfiability problems separately. In this paper we also
show that the infinite satisfiability reduces to the finite
satisfiability for the prenex fragment of $\seplogk{k}$.

For space reasons, some proofs are given in the extended technical
report \cite{EIP18b}. 

\section{Preliminaries}

In this section, we briefly review some usual definitions and notations.
We denote by $\zed$ the set of integers and by $\nat$ the set of
positive integers including zero. We define $\zed_{\infty} = \zed \cup
\set{\infty}$ and $\nat_{\infty} = \nat \cup \set{\infty}$, where for
each $n \in \zed$ we have $n + \infty = \infty$ and $n<\infty$. For a
countable set $S$ we denote by $\card{S} \in \nat_\infty$ the
cardinality of $S$. A decision problem is in
$\mathsf{(N)}\bigospace{n}$ if it can be decided by a
(nondeterministic) Turing machine in space $\bigO(n)$ and in $\pspace$
if it is in $\bigospace{n^c}$ for some input independent integer $c
\geq 1$.

\subsection{First Order Logic}

Let $\vars$ be a countable set of variables, denoted as $x,y,z$ and
$U$ be a sort. A \emph{function symbol} $f$ has $\#(f) \geq 0$
arguments of sort $U$ and a sort $\sigma(f)$, which is either the
boolean sort $\Bool$ or $U$. If $\#(f)=0$, we call $f$ a
\emph{constant}. We use $\bot$ and $\top$ for the boolean constants
false and true, respectively. First-order ($\fol$) terms $t$ and
formulae $\varphi$ are defined by the following grammar:
\[\begin{array}{rclcrcl}
t & := & x \mid f(t_1,\ldots,t_{\#(f)}) && \varphi & := & \bot \mid
\top \mid \varphi_1 \wedge \varphi_2 \mid \neg \varphi_1 \mid \exists
x ~.~ \varphi_1 \mid t_1 \teq t_2 \mid p(t_1,\ldots,t_{\#(p)})
\end{array}\]
where $x\in \vars$, $f$ and $p$ are function symbols, $\sigma(f)=U$
and $\sigma(p)=\Bool$. We write $\varphi_1 \vee \varphi_2$ for
$\neg(\neg\varphi_1 \wedge \neg\varphi_2)$, $\varphi_1 \rightarrow
\varphi_2$ for $\neg\varphi_1 \vee \varphi_2$, $\varphi_1
\leftrightarrow \varphi_2$ for $\varphi_1 \rightarrow \varphi_2 \wedge
\varphi_2 \rightarrow \varphi_1$ and $\forall x ~.~ \varphi$ for
$\neg\exists x ~.~ \neg\varphi$. The size of a formula $\varphi$,
denoted as $\size{\varphi}$, is the number of occurrences of symbols
needed to write it down. Let $\fv{\varphi}$ be the set of variables
that occur free in $\varphi$, i.e.\ not in the scope of a quantifier.

First-order formulae are interpreted over $\fol$-structures (called
structures, when no confusion arises) $\astruct =
(\U,\astore,\afunc)$, where $\U$ is a countable set, called the
\emph{universe}, the elements of which are called \emph{locations},
$\astore : \vars \rightharpoonup \U$ is a mapping of variables to
locations, called a \emph{store} and $\afunc$ interprets each function
symbol $f$ by a function $f^\afunc : \U^{\#(f)} \rightarrow \U$, if
$\sigma(f) = U$ and $f^\afunc : \U^{\#(f)} \rightarrow
\{\bot^\afunc,\top^\afunc\}$ if $\sigma(f)= \Bool$, with $\bot^\afunc
\not = \top^\afunc$. A structure $(\U,\astore,\afunc)$ is
\emph{finite} when $\card{\U} \in \nat$ and \emph{infinite} otherwise.

We write $\astruct \models \varphi$ iff $\varphi$ is true when
interpreted in $\astruct$. This relation is defined recursively on the
structure of $\varphi$, as usual. When $\astruct \models \varphi$, we
say that $\astruct$ is a \emph{model} of $\varphi$.  A formula is
\emph{satisfiable} when it has a model. We write $\varphi_1 \models
\varphi_2$ when every model of $\varphi_1$ is also a model of
$\varphi_2$ and by $\varphi_1 \equiv \varphi_2$ we mean $\varphi_1
\models \varphi_2$ and $\varphi_1 \models \varphi_2$.  The
\emph{(in)finite satisfiability problem} asks, given a formula
$\varphi$, whether a (in)finite model exists for this formula.

The Bernays-Sch\"onfinkel-Ramsey fragment of $\fol$ [$\bsr(\fol)$] is
the set of sentences $\exists x_1 \ldots \exists x_n \forall y_1
\ldots \forall y_m ~.~ \varphi$, where $\varphi$ is a quantifier-free
formula in which all function symbols $f$ of arity $\#(f)>0$ have sort
$\sigma(f)=\Bool$. 

\subsection{Separation Logic}

Let $k \in \nat$ be a strictly positive integer. The logic $\seplogk{k}$
is the set of formulae generated by the grammar: 
\[\begin{array}{rcl}
\varphi & := & \bot \mid \top \mid \emp \mid x \teq y \mid x \mapsto
(y_1,\ldots,y_k) \mid \varphi \wedge \varphi \mid \neg \varphi \mid
\varphi * \varphi \mid \varphi \wand \varphi \mid \exists x ~.~
\varphi
\end{array}\]
where $x,y,y_1,\ldots,y_k \in \vars$. The connectives $*$ and $\wand$
are respectively called the \emph{separating conjunction} and
\emph{separating implication} (\emph{magic wand}). We denote by
$\vec{y}$ the tuple $(y_1,\ldots,y_k) \in \vars^k$. The size of an
$\seplogk{k}$ formula $\varphi$, denoted $\size{\varphi}$, is the
number of symbols needed to write it down.

$\seplogk{k}$ formulae are interpreted over
$\seplog$-\emph{structures} (called structures when no confusion
arises) $\I = (\U, \astore, \aheap)$, where $\U$ and $\astore$ are as
before and $\aheap : \U \rightharpoonup_{\mathit{fin}} \U^k$ is a
finite partial mapping of locations to $k$-tuples of locations, called
a \emph{heap}. As before, a structure $(\U,\astore,\aheap)$ is finite when $\card{\U}
\in \nat$ and infinite otherwise. 

Given a heap $\aheap$, we denote by $\dom(\aheap)$ the domain of the
heap, by \(\img(\aheap) \isdef \{ \ell_i \mid \exists \ell \in
\dom(\aheap), \aheap(\ell) = (\ell_1,\dots,\ell_k), i \in [1,k]\}\)
and by \(\elemh{\aheap} = \dom(\aheap) \cup \img(\aheap)\) the set of
elements either in the domain or the image of the heap. For a store
$\astore$, we define $\img(\astore) \isdef \set{\ell \mid x \in
  \vars,~ \astore(x)=\ell}$. Two heaps $\aheap_1$ and $\aheap_2$ are
\emph{disjoint} iff $\dom(\aheap_1) \cap \dom(\aheap_2) = \emptyset$,
in which case $\aheap_1 \uplus \aheap_2$ denotes their union, where
$\uplus$ is undefined for non-disjoint heaps. The relation
$(\U,\astore,\aheap) \models \varphi$ is defined inductively, as
follows:
\[\begin{array}{lcl}
(\U,\astore,\aheap) \models \emp & \iff & \aheap = \emptyset \\ 
(\U,\astore,\aheap) \models x \mapsto (y_1,\ldots,y_k) & \iff &
\aheap = \set{\tuple{\astore(x),(\astore(y_1), \ldots, \astore(y_k))}}  \\
(\U,\astore,\aheap) \models \varphi_1 * \varphi_2 & \iff & 
\text{there exist disjoint heaps}\ \aheap_1,\aheap_2 
\text{ such that } \aheap=\aheap_1\uplus \aheap_2 \\
&& \text{and } (\U,\astore,\aheap_i) \models \varphi_i \text{, for $i = 1,2$} \\
(\U,\astore,\aheap) \models \varphi_1 \wand \varphi_2 & \iff &
\text{for all heaps $\aheap'$ disjoint from $\aheap$}
\text{ such that } (\U,\astore,\aheap') \models \varphi_1 \text{,} \\
&& \text{we have } (\U,\astore,\aheap'\uplus\aheap) \models \varphi_2
\end{array}\]
The semantics of equality, boolean and first-order connectives is the
usual one. Satisfiability, entailment and equivalence are defined for
$\seplogk{k}$ as for $\fol$ formulae. The (in)finite satisfiability
problem for $\seplogk{k}$ asks whether a (in)finite model exists for a
given formula. We write $\phi \equiv^{\fincard} \psi$ [$\phi
  \equiv^{\infcard} \psi$] whenever $(\U,\astore,\aheap) \models \phi
\iff (\U,\astore,\aheap) \models \psi$ for every finite
     [infinite] structure $(\U,\astore,\aheap)$.

The prenex fragment of $\seplogk{k}$ [$\prenex(\seplogk{k})$] is the
set of sentences $Q_1 x_1 \ldots Q_n x_n ~.~ \phi$, where $Q_1,
\ldots, Q_n \in \set{\exists, \forall}$ and $\phi$ is a
quantifier-free $\seplogk{k}$ formula. Unlike $\fol$, where each
formula is equivalent to a linear-size formula in prenex form, there
are $\seplogk{k}$ formulae that do not have a prenex form
equivalent. For instance, $\phi * \forall x ~.~ \psi(x) \not\equiv
\forall x ~.~ \phi * \psi(x)$ and dually, $\phi \wand \exists x ~.~
\psi(x) \not\equiv \exists x ~.~ \phi \wand \psi(x)$, where $\phi$ and
$\psi$ are arbitrary $\seplogk{k}$ formulae. 

The Bernays-Sch\"onfinkel-Ramsey fragment of $\seplogk{k}$
[$\bsr(\seplogk{k})$] is the set of sentences $\exists x_1 \ldots
\exists x_n \forall y_1 \ldots \forall y_m ~.~ \phi$, where $\phi$ is
a quantifier-free $\seplogk{k}$ formula. Since there are no function
symbols of arity greater than zero in $\seplogk{k}$, there are no
restrictions, other than the form of the quantifier prefix, defining
$\bsr(\seplogk{k})$.

\subsection{Test Formulae for $\seplogk{k}$}

This section contains a number of definitions and results from
\cite{EIP18}, needed for self-containment. For more details, the
interested reader is pointed towards \cite{EIP18}.
\begin{definition}\label{def:test-formulae}
The following patterns are called \emph{test formulae}:
\[\begin{array}{c}
\begin{array}{rclcrcl}
x \pto \vec{y} & \defequal & x \mapsto \vec{y} * \top 
&\quad\quad& \len{U} \geq n & \defequal & \top \septraction \len{h} \geq n,~ n \in \nat \\
\alloc(x) & \defequal & x \mapsto \underbrace{(x,\ldots,x)}_{k \text{ times}} \wand \bot 
&& \len{h} \geq \len{U} - n & \defequal & \len{h} \geq n + 1 \wand \bot, n \in \nat \\
\end{array} \\
\begin{array}{rcl}
\len{h} \geq n & \defequal & \left\{\begin{array}{ll}
\len{h} \geq n-1 * \neg\emp, & \text{if $n>0$} \\
\top, & \text{if $n = 0$} \\
\bot, & \text{if $n = \infty$}
\end{array}\right. 
\end{array}
\end{array}\]
and $x \teq y$, where $x,y \in \vars$, $\vec{y} \in \vars^k$ and $n
\in \nat_\infty$ is a positive integer or $\infty$. 
\end{definition}
The test formulae of the form $\len{U}\geq n$ and
$\len{h}\geq\len{U}-n$ are called \emph{domain dependent} and the rest
\emph{domain independent}. A \emph{literal} is a test formula or its
negation. 

The semantics of test formulae is intuitive: $x \pto \vec{y}$ holds
when $x$ denotes a location and $\vec{y}$ is the image of that
location in the heap, $\alloc(x)$ holds when $x$ denotes a location in
the domain of the heap (allocated), $\len{h}\geq n$, $\len{U}\geq n$
and $\len{h}\geq\len{U}-n$ are cardinality constraints involving the
size of the heap, denoted $\len{h}$ and that of the universe, denoted
$\len{U}$. We recall that $\len{h}$ ranges over $\nat$, whereas
$\len{U}$ is always interpreted as a number larger than $\len{h}$ and
possibly infinite.

Observe that not all atoms of $\seplogk{k}$ are test formulae, for
instance $x \mapsto \vec{y}$ and $\emp$ are not test
formulae. However, we have the equivalences $x \mapsto \vec{y} \equiv
x \pto \vec{y} \wedge \neg\len{h} \geq 2$ and $\emp \equiv \neg\len{h}
\geq 1$. Moreover, for any $n \in \nat$, the test formulae
$\len{U}\geq n$ and $\len{h}\geq \len{U}-n$ become trivially true and
false, respectively, if we consider the universe to be infinite.

The following result establishes a translation of quantifier-free
$\seplogk{k}$ formulae into boolean combinations of test
formulae. This translation relies on the notion of a minterm. 

\begin{definition}\label{def:minterm}
A \emph{minterm} $M$ is a set (conjunction) of literals containing:
\begin{compactitem}
\item exactly one literal $\len{h} \geq \minheap_M$ and one literal
  $\len{h} < \maxheap_M$, where $\minheap_M \in \nat \cup \set{\len{U}
  - n \mid n \in \nat}$ and $\maxheap_M \in \nat_{\infty} \cup
  \set{\len{U} - n \mid n \in \nat}$, and
\item exactly one literal of the form $\len{U} \geq n$ and at most one
  literal of the form $\len{U} < n$.
\end{compactitem}
\end{definition}

One of the results in \cite{EIP18} is that, for each quantifier-free
$\seplogk{k}$ formula $\phi$, it is possible to define a disjunction
on minterms that preserves the finite models of $\phi$.  We denote the
set of minterms in the disjunction as $\finmt{\phi}$, where
$\finmt{.}$ is an effectively computable function, defined recursively
on the structure of $\phi$.

\begin{lemma}\label{lemma:sl-mnf}
  Given a quantifier-free $\seplogk{k}$ formula $\phi$, $\finmt{\phi}$
  is a finite set of minterms and we have $\phi \equivfin \bigvee_{M
    \in \finmt{\phi}} M$.
\end{lemma}
\proof{See \cite[Lemma 5]{EIP18}. \qed}

Given a quantifier-free $\seplogk{k}$ formula $\phi$, the number of
minterms occurring in $\finmt{\phi}$ is exponential in the size of
$\phi$, in the worst case. Therefore, an optimal decision procedure
cannot generate and store these sets explicitly, but rather must
enumerate minterms lazily.  The next lemma shows that it is possible
to check whether $M \in \finmt{\phi}$ using space bounded by a
polynomial in $\size{\phi}$.  For a boolean combination of test
formulae $\phi$, we denote by $\maxn{\phi}$ the maximum $n \in \nat$
that occurs in an atom of the form $\len{h} \geq n$ or $\len{U} \geq
n$ in $\phi$.

\begin{lemma}\label{lemma:pspacemt}
For every $\seplogk{k}$ formula $\phi$, the size of every minterm $\finmt{\phi}$ 
is polynomial w.r.t.\ $\size{\phi}$. In particular, $\max_{M\in \finmt{\phi}} \maxn{M}$ is polynomial 
w.r.t.\ $\size{\phi}$.
Furthermore, %Given a minterm $M$ and an $\seplogk{k}$ formula $\phi$, 
given a minterm $M$, the problem
of checking whether $M \in \finmt{\phi}$ is in \pspace.
\end{lemma}
\proof{See \cite[Lemma 8 and Corollary 1]{EIP18}. \qed}

\section{The $\prenex(\seplogk{1})$ Fragment is Decidable}

The first result of this paper is the decidability of the prenex
fragment of $\seplogk{1}$. In particular, this shows that
$\prenex(\seplogk{k})$ is strictly less expressive than $\seplogk{k}$,
because $\seplogk{1}$ has been shown to be at least as expressive as
Second Order Logic, thus having an undecidable satisfiability problem
\cite[Theorem 6.11]{BrocheninDemriLozes11}.

\subsection{From Infinite to Finite Satisfiability}

We begin by showing that the infinite satisfiability problem can be
reduced to the finite satisfiability problem for prenex
$\seplog$-formulae. The intuition is that two $\seplog$-structures
defined on the same heap and store can be considered equivalent if
both have enough locations outside of the heap.
\begin{definition}
\label{def:equiv}
Let $X$ be a set of variables and let $n \in \nat$.  Two
$\seplog$-structures $\I = (\U,\astore,\aheap)$ and $\I' =
(\U',\astore',\aheap')$ are {\em \equivalent{X}{n}} (written $\I
\equivs{X}{n} \I'$) iff the following conditions hold:
\begin{enumerate}
\item{$\aheap = \aheap'$.
\label{equiv:h}
}
\item{ For all $x,y \in X$, $\astore(x)=\astore(y) \iff
  \astore'(x)=\astore'(y)$.
\label{equiv:X}
}
\item{For every $x \in X$, if $\astore(x)\in \elemh{\aheap}$ or
  $\astore'(x)\in \elemh{\aheap}$ then $\astore(x)=\astore'(x)$.
\label{equiv:x}
}
\item{$\card{\U \setminus \elemh{\aheap}} \geq n+\card{X}$ and
  $\card{\U' \setminus \elemh{\aheap}} \geq n+\card{X}$.}
 \label{equiv:n}
\end{enumerate}
\end{definition}
Note that Condition \ref{equiv:h} entails that $\elemh{\aheap}
\subseteq \U \cap \U'$.  Next, we prove that any two
$\seplog$-structures that are \equivalent{\fv{\phi}}{m}\ are also
indistinguishable by any formula $\phi$ prefixed by $m$ quantifiers.

\begin{proposition}
\label{prop:equivs}
Let $\phi = \Qu_1 x_1 \dots \Qu_m x_m~.~ \psi$ be a prenex
$\seplogk{k}$ formula, with $\Qu_i \in \{ \forall,\exists \}$ for $i =
1,\dots,m$.  Assume that $\psi$ is a quantifier-free boolean
combination of domain independent test formulae. If $\I
\equivs{\var{\phi}}{m} \I'$ and $\I \models \phi$ then $\I' \models
\phi$.
\end{proposition}
\ifLongVersion
\proof{
Let $\I = (\U, \astore,\aheap)$ and $\I' = (\U',\astore',\aheap')$.
Assume that $\I \equivs{\var{\phi}}{m} \I'$.  By Condition
\ref{equiv:h} in Definition \ref{def:equiv} we have $\aheap =
\aheap'$. The proof is by induction on $m$.
\begin{compactitem}
\item{If $m = 0$, we have $\phi = \psi$, we show that $\I$ and $\I'$
  agree on every atomic formula in $\phi$, which entails by an
  immediate induction that they agree on $\phi$.  By Condition
  \ref{equiv:X} in Definition \ref{def:equiv}, we already know that
  $\I$ and $\I'$ agree on every atom $x \teq x'$ with $x,x' \in
  \var{\phi}$. By Condition \ref{equiv:h}, $\I$ and $\I'$ agree on all
  atoms $\len{h} \geq n$. Consider an atom $\ell \in \{ y_0 \pto
  (y_1,\dots,y_k), \alloc(y_0) \}$, with $y_0,\dots,y_k \in
  \var{\phi}$.  If for every $i \in \interv{0}{k}$ we have
  $\astore(y_i) \in \elemh{\aheap}$ then by Condition \ref{equiv:x} we
  deduce that $\astore'$ and $\astore$ coincide on $y_0,\dots,y_k$
  hence $\I$ and $\I'$ agree on $\ell$ because they share the same
  heap.  The same holds if $\astore'(y_i) \in \elemh{\aheap}$,
  $\forall i \in \interv{0}{k}$.  If both conditions are false, then
  we must have $\I \not \models \ell$ and $\I' \not \models \ell$, by
  definition of $\elemh{\aheap}$, thus $\I$ and $\I'$ also agree on
  $\ell$ in this case.  }

\item{ Assume that $m \geq 1$ and $Q_1 = \exists$. Then $\phi =
  \exists x_1~.~\phi'$.  Assume that $\I \models \phi$.  Then there
  exists $e \in \U$ such that $(\U,\astore[x_1 \mapsto e],\aheap)
  \models \phi'$.  We construct an element $e' \in \U'$ as follows.
  If $e = \astore(y)$, for some $y \in \var{\phi}$, then we let $e' =
  \astore'(y)$. If $\forall y \in \var{\phi}, e \not = \astore(y)$ and
  if $e \in \elemh{\aheap}$ then we let $e' = e$.  Otherwise, $e'$ is
  an arbitrarily chosen element in $\U' \setminus
  (\astore'(\var{\phi}) \cup \elemh{\aheap})$. Such an element
  necessarily exists, because by Condition \ref{equiv:n} in Definition
  \ref{def:equiv}, $\U'$ contains at least $m + \card{\var{\phi}} \geq
  1+\card{\astore(\var{\phi})}$ elements distinct from those in
  $\elemh{\aheap}$.  Let $\J = (\U,\astore[x_1 \mapsto e],\aheap)$ and
  $\J' = (\U,\astore[x_1 \mapsto e],\aheap)$.  We now prove that $\J
  \equivs{\var{\phi} \cup \{ x_1\}}{m-1} \J'$. This entails the
  desired results since by the induction hypothesis we deduce $\J'
  \models \phi'$, hence $\I' \models \phi$.
   
   Condition \ref{equiv:h} trivially holds. For Condition
   \ref{equiv:x}, assume that there exists a variable $x$ such that
   $\astore[x_1 \mapsto e](x)\in \elemh{\aheap}$ or $\astore'[x_1
     \mapsto e'](x)\in \elemh{\aheap}$ and $\astore[x_1 \mapsto
     e](x)\not =\astore'[x_1 \mapsto e'](x)$. Since $\I
   \equivs{\var{\phi}}{m} \I'$, we must have $\astore(x)\in
   \elemh{\aheap} \vee \astore'(x)\in \elemh{\aheap} \Rightarrow
   \astore(x) =\astore'(x)$ if $x \in \var{\phi}$, thus necessarily $x
   = x_1$, hence $\astore[x_1 \mapsto e](x_1)=e$ and $\astore'[x_1
     \mapsto e'](x_1) = e'$.  If $e' = \astore'(y)$, for some $y \in
   \var{\phi}$ such that $\astore(y)=e$, then the proof follows from
   the fact that $\astore(y)\in \elemh{\aheap} \vee \astore'(y)\in
   \elemh{\aheap} \Rightarrow \astore(y) =\astore'(y)$, because $\I
   \equivs{\var{\phi}}{m} \I'$ and $y \in \var{\phi}$.  If the
   previous condition does not hold and $e \in \elemh{\aheap}$ then we
   must have $e' = e$, by definition of $e'$, which contradicts our
   hypotheses.  Otherwise, it cannot be the case that $e' \in
   \elemh{\aheap}$, by definition of $e'$, thus the disjunction $e \in
   \elemh{\aheap} \vee e' \in \elemh{\aheap}$ cannot hold.

   Condition \ref{equiv:n} follows from the fact that $\I
   \equivs{\var{\phi}}{m} \I'$ because we have $m-1+\card{\var{\phi}
     \cup \{ x_1 \}} = m + \card{\var{\phi}}$.
   
   We now establish Condition \ref{equiv:X}. Let $x,x' \in \var{\phi}
   \cup \{ x_1 \}$. If $x,x' \in \var{\phi}$ then $\astore[x_1 \mapsto
     e]$ and $\astore'[x_1 \mapsto e']$ coincide with $\astore$ and
   $\astore'$ respectively on $x$ and $x'$, hence $\J$ and $\J'$ must
   agree on $x \teq x'$ since $\I \equivs{\var{\phi}}{m} \I'$.
   Otherwise, we may assume, w.l.o.g., that $x = x_1$ and $x' \not =
   x_1$ (the proof for the case where $x = x'$ is immediate).  If $e =
   \astore(y)$, for some $y \in \var{\phi}$, then $\J \models x \teq
   x' \iff \I \models y \teq x'$.  By definition of $e'$, we also have
   $e'=\astore'(y)$ thus $\J' \models x \teq x' \iff \I' \models y
   \teq x'$.  Since $\I \equivs{\var{\phi}}{m} \I'$ and $y,x' \in
   \var{\phi}$, we must have $\I \models y \teq x' \iff \I' \models y
   \teq x'$ thus the proof is completed.  If the previous condition
   does not hold then necessarily $e \not = \astore(x')$, and thus $\J
   \not \models x_1 \teq x'$.  If $e \in \elemh{\aheap}$, then by
   definition of $e'$, $e' = e$.  If $\J' \models x_1 \teq x'$ then we
   must have $\astore'(x') = \astore'(x_1) = e' = e \in
   \elemh{\aheap}$, which by Condition \ref{equiv:x} entails that
   $\astore'(x') = \astore(x') = e$, hence $\J \models x_1 \teq x'$, a
   contradiction.  Finally, if $e \not \in \elemh{\aheap}$, then by
   definition of $e'$, $e'$ cannot occur in $\astore'(\var{\phi})$,
   thus $\J \not \models x_1 \teq x'$.}

\item{Finally, assume that $m \geq 1$ and $Q_1 = \forall$. Then $\phi
  = \forall x_1~.~\phi'$. Assume that $\I \models \phi$. Let $\phi_2 =
  \exists x_1~.~\phi'_1$, where $\phi'_1$ denotes the nnf of $\neg
  \phi'$. Assume that $\I' \not \models \phi$, then $\I' \models
  \phi_2$, because $\neg \phi \equiv \exists x_1~.~ \neg \phi' \equiv
  \exists x_1~.~ \phi'_1 = \phi_2$.  By the previous case, using the
  symmetry of $\equivs{\var{\phi}}{m}$ and the fact that $\phi$ and
  $\phi_2$ have exactly the same free variables and number of
  quantifiers, we know that $\I \models \phi_2$, i.e. $\I \not \models
  \phi$, a contradiction. \qed}
\end{compactitem}
}
\fi

The formulas $x \in h$ and $\alldistinct(x_1,\dots,x_n)$ are
shorthands for the formulas $\exists y_0,y_1,\dots y_k~.~(y_0 \pto
(y_1,\dots,y_k) \wedge \bigvee_{i=0}^k x \teq y_i)$ and
$\bigwedge_{i=1}^n \bigwedge_{j=1}^{i-1} \neg (x_i \teq x_j)$,
respectively. We define the formula: \[\form{p} \defequal \exists
x_1,\dots,x_p~.~ (\alldistinct(x_1,\dots,x_p) \wedge \bigwedge_{i=1}^p
\neg x_i \in h)\] It is clear that $(\U,\astore,\aheap) \models
\form{p}$ iff $\card{\U \setminus \elemh{\aheap}} \geq p$.  In
particular, $\form{p}$ is always true on infinite domains. Observe,
moreover, that $\lambda_p$ belongs to the $\prenex(\seplogk{k})$
fragment, for any $p\geq2$ and any $k\geq1$.

The following lemma reduces the infinite satisfiability problem to the
finite version of it. This is done by adding an axiom ensuring that
there are enough locations outside of the heap. Note that there is no
need to consider test formulae of the form $\len{U} \geq n$ and
$\len{h} \geq \len{U} - n$ because they alway evaluate to true and,
respectively, false, on infinite $\seplog$-structures.

\begin{lemma}\label{lem:infinitetofinite}
Let $\phi = \Qu_1 x_1 \dots \Qu_m x_m~.~ \psi$ be a prenex
$\seplogk{k}$ formula, where $\Qu_i \in \{ \forall,\exists \}$ for $i
= 1,\dots,m$ and $\var{\phi} = \emptyset$.  Assume that $\psi$ is a
boolean combination of test formulas of the form $x \teq y$ or $x \pto
(y_1,\dots,y_k)$ or $\alloc(x)$ or $\len{h} \geq n$. The two following
assertions are equivalent.
\begin{compactenum}
\item{$\phi$ admits an infinite model.}
\item{
  $\phi \wedge \form{m}$ admits a finite model.}
\end{compactenum} 
\end{lemma}
\proof{
  $(1) \Rightarrow (2)$: Assume that $\phi$ admits an infinite
    model $(\U,\astore,\aheap)$.  Let $\U'$ be a finite subset of $\U$
    including $\elemh{\aheap}$ plus $m$ additional elements.  It is
    clear that $(\U,\astore,\aheap) \equivs{\emptyset}{m}
    (\U',\astore,\aheap)$.  Indeed, Condition \ref{equiv:h} holds
    since the two structures share the same heap, Conditions
    \ref{equiv:X} and \ref{equiv:x} trivially hold since the
    considered set of variables is empty, and Condition \ref{equiv:n}
    holds since $\U$ is infinite and the additional elements in $\U'$
    do not occur in $\elemh{\aheap}$.  Thus $(\U',\astore,\aheap)
    \models \phi$ by Proposition \ref{prop:equivs}.  Furthermore,
    $(\U',\astore,\aheap) \models \form{m}$, by definition of $\U'$.
 
    $(2) \Rightarrow (1)$: Assume that 
    $\phi \wedge \form{m}$ has a finite model $(\U,\astore,\aheap)$. 
    Let $\U'$ be any infinite set containing $\U$. 
    Again, we have $(\U,\astore,\aheap) \equivs{\emptyset}{m} (\U',\astore,\aheap)$.
    As in the previous case, Conditions \ref{equiv:h}, \ref{equiv:X} and \ref{equiv:x} trivially hold, and 
    Condition \ref{equiv:n} holds since $\U'$ is infinite and $(\U,\astore,\aheap) \models  \form{m}$.
   By Proposition \ref{prop:equivs}, we deduce that 
   $(\U',\astore,\aheap) \models \phi$. \qed
}

\subsection{Translating $\prenex(\seplogk{1})$ into First-Order Logic}

After reduction of the infinite to the finite satisfiability problem,
the decidability of the latter for $\prenex(\seplogk{1})$ is
established by reduction to the finite satisfiability of the
$[\mathit{all},(\omega), (1)]_=$ fragment of $\fol$, with an arbitrary
number of monadic boolean function symbols and one function symbol $f$
of sort $\sigma(f)=U$. The decidability of this fragment is a
consequence of the celebrated Rabin's Tree Theorem, which established
the decidability of the monadic theory of the infinite binary tree
\cite{Rabin69}.

In the following, we define an equivalence-preserving (on finite
structures) translation of $\seplogk{k}$ into $\fol$. Let $\domh$ be a
unary predicate symbol and let $\funch_i$ (for $i = 1,\dots,k$) be
unary function symbols. We define the following transformation from
quantified boolean combinations of test formulae into first order
formulae:
\[
\begin{array}{rcl}
  \SLtoFO{x \teq y} & \defequal & x \teq y \\
  \SLtoFO{x \pto (y_1,\dots,y_k)} & \defequal & \domh(x) \wedge \bigwedge_{i=1}^k y_i \teq \funch_i(x) \\
  \SLtoFO{\alloc(x)} & \defequal & \domh(x) \\
  \SLtoFO{\neg \phi} & \defequal & \neg \SLtoFO{\phi} \\
  \SLtoFO{\phi_1 \bullet \phi_2} & \defequal & 
  \SLtoFO{\phi_1} \bullet \SLtoFO{\phi_2}  \qquad \text{if } \bullet \in \{ \wedge, \vee, \rightarrow, \leftrightarrow \} \\
  \SLtoFO{\Qu x~.~ \phi} & \defequal & \Qu x ~.~ \SLtoFO{\phi} \qquad \text{if }  \Qu \in \{ \exists, \forall \} \\
  \SLtoFO{\len{U} \geq n} & \defequal &	\exists x_1,\dots,x_n~.~ \alldistinct(x_1,\dots,x_n) \\
  \SLtoFO{\len{h} \geq n} & \defequal &	\exists x_1,\dots,x_n~.~  \alldistinct(x_1,\dots,x_n) \wedge \bigwedge_{i=1}^n \domh(x_i) \\
  \SLtoFO{\len{h} \geq \len{U}-n} & \defequal &	\exists x_1,\dots,x_n\forall y~.~ \bigwedge_{i=1}^n y \not \teq x_i \rightarrow \domh(y) \\
\end{array}
\]

\begin{proposition}\label{prop:red}
  Let $\phi$ be a quantified boolean combination of test formulae.
  The formula $\phi$ has a finite $\seplog$ model iff $\SLtoFO{\phi}$
  has a finite $\fol$ model.
\end{proposition}
\proof{ A $\fol$-structure $\I = (\U, \astore, \afunc)$ on the
  signature $\domh$, $\funch_1$,\dots,$\funch_k$ {\em corresponds} to
  an $\seplog$-structure $\I' = (\U',\astore',\aheap)$ iff $\U= \U'$,
  $\astore = \astore'$, $\domh^\afunc = \dom(\aheap)$ and for every $j
  \in \interv{1}{k}$, $\funch_j^\afunc(x) = y_j$ if $\aheap(x) =
  (y_1,\dots,y_k)$.  It is clear that for every finite first-order
  structure $\I$ there exists a finite $\seplog$-structure $\I'$ such
  that $\I$ corresponds to $\I'$ and vice-versa.  Furthermore, if $\I$
  corresponds to $\I'$ then it is straightforward to check that $\I'
  \models \phi \iff \I \models \SLtoFO{\phi}$.\qed }

Given a formula $\psi = Q_1 x_1 \ldots Q_n x_n ~.~ \phi$ of
$\prenex(\seplogk{1})$, where $\phi$ is a quantifier-free
$\seplogk{1}$ formula, consider the expansion of $\phi$ as a
disjunction of minterms $\mu \isdef \bigvee_{M \in \finmt{\phi}}
M$. By Lemma \ref{lemma:sl-mnf}, we have $\phi \equivfin \mu$, thus
$\psi \equivfin Q_1 x_1 \ldots Q_n x_n ~.~ \mu$. By Proposition
\ref{prop:red}, $\psi$ has a finite $\seplog$ model iff $\SLtoFO{Q_n
  x_n ~.~ \mu}$ has a finite $\fol$ model. Moreover, it is easy to see
that $\SLtoFO{Q_n x_n ~.~ \mu}$ belongs to the $[\mathit{all},
  (\omega), (1)]_=$ fragment of $\fol$, whose finite satisfiability
problem is decidable \cite[Corollary
  7.2.12]{BorgerGradelGurevich97}. The following theorem summarizes
the result:

\begin{theorem}\label{thm:seplog-dec}
  The finite and infinite satisfiability problems are decidable for
  $\prenex(\seplogk{1})$.
\end{theorem}
\ifLongVersion
\proof{Using Lemma \ref{lem:infinitetofinite}, Proposition
  \ref{prop:red} and \cite[Corollary
    7.2.12]{BorgerGradelGurevich97}. \qed}
\fi

\section{The $\prenex(\seplogk{1})$ Fragment is not Elementary Recursive}

This section is concerned with the computational complexity of the
(in)finite satisfiability problem(s) for the $\prenex(\seplogk{1})$
fragment. We use the fact that the $[\mathit{all}, (\omega), (1)]_=$
fragment of $\fol$ is nonelementary and obtain a similar lower bound
by an opposite reduction, from the satisfiabilty of $[\mathit{all},
  (\omega), (1)]_=$ to that of $\prenex(\seplogk{1})$. This reduction,
in the finite and infinite case, respectivelly, is carried out by the
following propositions: 

\begin{proposition}\label{prop:fin}
  There is a polynomial reduction of the finite satisfiability problem
  for $\fol$ formulae with one monadic function symbol to the finite
  satisfiability problem for $\prenex(\seplogk{1})$ formulae.
\end{proposition}
\proof{ The reduction is immediate: it suffices to add the axiom:
  $\forall x~.~ \alloc(x)$ (i.e., the heap is total) and replace all
  equations of the form $f(x) \teq y$ by $x \pto y$ (by flattening we
  may assume that all the equations occurring in the formula are of
  the form $f(x) \teq y$ or $x \teq y$, where $x,y$ are variables). It
  is straightforward to check that satisfiability is preserved. \qed}

\begin{proposition}\label{prop:infin}
  There is a polynomial reduction of the finite satisfiability problem
  for $\fol$ formulae with one monadic function symbol to the infinite
  satisfiability problem for $\prenex(\seplogk{1})$ formulae.
\end{proposition}
\proof{ We may apply the same transformation as above on equations
  $f(x) \teq y$, but this time the axiom $\forall x~.~ \alloc(x)$
  cannot be added as it would make the resulting formula
  unsatisfiable. Instead, we add the axiom $\neg\emp \wedge \forall
  x,y~.~ x \pto y \rightarrow \alloc(y)$, and we replace every
  quantification $\forall x~.~ \phi$ (resp.\ $\exists x~.~ \phi$) by a
  quantification over the domain of the heap: $\forall x~.~ \alloc(x)
  \rightarrow \phi$ (resp.\ $\exists x~.~ \alloc(x) \wedge \phi$).
  Again, it is straightforward to check that satisfiability is
  preserved. Note that infinite satisfiability is equivalent to finite
  satisfiability here since the quantifications range over elements
  occurring in the heap. The domain of the (finite) first-order
  interpretation is encoded as the domain of the heap. \qed }

The main difficulty here is the lack of a direct result stating that
the \emph{finite} satisfiability problem for $[\mathit{all}, (\omega),
  (1)]_=$ is nonelementary. Instead the result of \cite[Theorem
  7.2.15]{BorgerGradelGurevich97} considers arbitrary $\fol$
structures, in which the cardinality of the universe is not
necessarily finite. In the following we show that this result can be
strenghtened to considering finite structures only. Observe that this
is not automatically the case for $\fol$ formulae with one monadic
function symbol, for instance, the formula \(\exists x\forall y~.~ x
\not \teq f(y) \wedge \forall y,z, ~.~ f(y)\teq f(z) \rightarrow y
\teq z \) is satisfiable only on infinite $\fol$ structures. However,
this is the case for the formula obtained in \cite[Theorem
  7.2.15]{BorgerGradelGurevich97} by reduction from domino the problem
of nonelementary size, defined below:

\begin{definition}\label{def:domino}
  A \emph{domino system} is a tuple $\mathcal{D} = (D,H,V)$, where $D$
  is a finite set of tiles and $H,V \subseteq D \times D$. For some
  $t\geq2$, let $Z_t \times Z_t$ be a \emph{torus}, where
  $Z_t=([0,t-1],\successor)$ and $\successor(n) \isdef (n+1)\mod t$,
  for all $n \in [0,t-1]$. We say that $\mathcal{D}$ \emph{tiles} $Z_t
  \times Z_t$ \emph{with initial condition} $d_0\ldots d_{m-1} \in
  D^m$ iff there exists a mapping $\tau : Z_t \times Z_t \rightarrow
  D$ such that, for all $(x,y) \in Z_t \times Z_t$, we have
  $H(\tau(x,y),\tau(\successor(x),y))$ and
  $V(\tau(x,y),\tau(x,\successor(y)))$, and moreover $\tau(i,0)=d_i$,
  for all $i\in[0,m-1]$.
\end{definition}
Given a tower of exponentials \(\tower(n) =
\underbrace{2^{2^{\ldots^2}}}_n\), the existence of a tiling of
$Z_{\tower(n)} \times Z_{\tower(n)}$ with a given initial condition is a
nonelementary recursive problem \cite[Theorem
  6.1.2]{BorgerGradelGurevich97}. For the sake of self-containment, we
describe the main ingredients of the reduction from this problem to
the satisfiability of $[\mathit{all},(\omega),(1)]_=$ on arbitrary
$\fol$-structures. 

Suppose that $D=\set{d_1,\ldots,d_r}$. First, we express the tiling
conditions (Definition \ref{def:domino}) by a formula $\theta$, using
$r+1$ binary boolean functions $P_0,\ldots,P_r$,
where:\begin{compactenum}
\item $P_0(x,y)$ encodes the successor relation $\successor(x)=y$,
\item $P_i(x,y)$ holds iff $\tau(x,y)=d_i$, for all $i\in[1,r]$,
\item the horizontal and vertical adjacency conditions $H$ and $V$ are
  respected, and
\item there is an element $x_0$ such that the points
  $(x_0,\successor^i(x_0))$ are labeled with $w_i$, for all
  $i\in[0,m-1]$.
\end{compactenum}

Next, we assume that the $\fol$-structures encoding the tiling are
models of the formula \(\alpha \isdef \exists x\forall y ~.~ f(x) \teq
x \wedge f^ {n+1}(y) \teq x\), which states that the domain can be
viewed as a tree of height at most $n+1$, where the (necessarily
unique) element assigned to the variable $x$ is the root of the tree,
and where $f$ maps every other node to its parent.

Intuitively, the domain $[0,\tower(n)-1]$ will be represented by the
direct sons of the root. The main problem is ensuring that the
universe $Z_{\tower(n)}$ has size (at most) $\tower(n)$. To this end
we define inductively the equivalence relations $E_0, \ldots, E_n$ as:
\begin{compactenum}
\item{all nodes are $E_0$-equivalent, and}
\item{for $m\geq1$, two nodes are $E_m$-equivalent if for every
  $E_{m-1}$-equivalence class $K$, either both nodes have no child
  in $K$ or both nodes have a child in $K$.}
\end{compactenum}
Then, in each model of $\alpha$, there are at most $\tower(m)$
$E_m$-equivalence classes, for each $m\geq0$: all elements are
$E_0$-equivalent and the index of $E_m$ is at most that of $E_{m-1}$
squared, for all $m\geq1$. This is because any two elements $x$ and
$y$ can be distinguished by $E_m$ only if they have a pair of children
$(x_1,x_2)$ and $(y_1,y_2)$ each, such that $[x_i]_{E_{m-1}} \neq
[y_i]_{E_{m-1}} $, for some $i=1,2$, where $[x]_E$ is the equivalence
class of $x$ w.r.t. $E$. Moreover, we have $E_{m-1} \subseteq E_m$,
for all $m\geq1$, therefore $E_n = \cap_{i=0}^n E_i$.

We consider formulae $\beta_m(x,y)$ stating that $x$ and $y$ have
height at most $m$ and are $E_m$-equivalent and a formula $\delta(x)$,
stating that $x$ is a child of the root (asserted by $\alpha$) with at
most one child in each $E_{n-1}$ equivalence class. Then let \(\gamma
\isdef \forall x,y~.~ \delta(x) \wedge \delta(y) \wedge \beta_n(x,y)
\rightarrow x \teq y\). In any model $(\U,\astore,\afunc) \models
\alpha \wedge \gamma$ there are at most $\tower(n)$ elements $a$ such
that $(\U,\astore[x\leftarrow a],\afunc) \models \delta(x)$, because
there is at most one element in each $E_n$-equivalence class and there
are at most $\tower(n)$ such classes.

It remains to encode the fact that an element $(x,y) \in Z_\tower(n)
\times Z_\tower(n)$ is labeled by the tile $d_i$, i.e.\ that
$P_i(x,y)$ holds in any model of $\theta$. Since we assumed that
$\delta(x) \wedge \delta(y)$ holds, $x$ and $y$ have at most one child
in each $E_{n-1}$ equivalence class, thus each element can be
distinguished by the tuple $(n_1, \ldots, n_s)$ of numbers of children
in each $E_{n-1}$ equivalence class $K_1, \ldots, K_s$. We encode
$P_i(x,y)$ by assuming the existence of a node $z$ with $g_i(j,k) = 2
+ 4i + 2j + k$ children in each class $K_1, \ldots, K_s$. This is
encoded by a formula $\pi_i(x,y)$. 

Finally, the $[\mathit{all}, (\omega), (1)]_=$ formula that states the
existence of a tiling of $Z_\tower(n) \times Z_\tower(n)$ is obtained
from $\theta$ by replacing each quantifier $\exists x ~.~ \phi$ by
$\exists x ~.~ \delta(x) \wedge \phi$ and $\forall x ~.~ \phi$ by
$\forall x ~.~ \delta(x) \rightarrow \phi$ and each occurrence of a
predicate symbol $P_i(x,y)$ by $\pi_i(x,y)$. 

\begin{lemma}\label{lem:book}
The finite satisfiability problem is not elementary recursive for
first order formulae built on a signature containing only one function
symbol of arity $1$ and the equality predicate.
\end{lemma}
\proof{
Let $\varphi$ be the formula encoding the existence of a tiling of
$Z_\tower(n) \times Z_\tower(n)$ by a tiling system
$\mathcal{D}=(D,H,V)$ and $\I = (\U,\astore,\afunc)$ be a model of
$\varphi$, with $\semf = f^{\afunc}$. We denote by $\rootnode$ the
root of the tree, i.e., the unique element of $\U$ with $(\U,
\astore[x \mapsto \rootnode], \afunc) \models \forall y ~.~ f(x) \teq
x \wedge f^ {n+1}(y) \teq x$.  Given $i\in [0,r]$ and $a,b \in \U$, if
$(\U, \astore[x \mapsto a,y \mapsto b], \afunc) \models \pi_i(x,y)$,
then we denote by $\nodesPi(i,a,b)$ a set containing an arbitrarily
chosen element $z$ satisfying $\propr(i,a,b)$ in the definition of
$\pi_i(x,y)$ along with all the children of $z$, otherwise
$\nodesPi(i,a,b)$ is empty. Observe that $\nodesPi(i,a,b)$ is always
finite because the number of children of $z$ in each equivalence class
is bounded by $g_i(1,1) = 2+4\times i+2+1\leq 2+4\times r+2+1$,
moreover the number of $E_n$-equivalence classes is finite.

We show that $\varphi$ admits a finite model $\I'$. The set $B$ of
elements $b$ such that $(\U, \astore[x \mapsto b],\afunc) \models
\delta(x)$ is finite. Let \(\Pi \isdef \bigcup \{ \nodesPi(i,a,b) \mid
a,b \in B, i \in [0,r] \}\). Since $B$ is finite and every set
$\nodesPi(i,a,b)$ is finite, $\Pi$ is also finite.  With each element
$a \in \U$ and each $E_n$-equivalence class $K$, we associate a set
$\nodesCh(a,K)$ containing exactly one child of $a$ in $K$ if such a
child exists, otherwise $\nodesCh(a,K)$ is empty. We now consider the
subset $\U'$ of $\U$ defined as the set of elements $a$ such that for
every $m \in \nat$, $\semf^m(a)$ occurs either in $\{ \rootnode \}
\cup B \cup \Pi$ or in a set $\nodesCh(b,K)$, where $b \in \U$ and $K$
is an $E$-equivalence class.  Note that $\rootnode \in \U'$ and that
if $a \in \U'$ then necessarily $\semf(a) \in \U'$.  Furthermore, if
$\semf(b) \in \U'$ and $b \in \nodesCh(\semf(b),K)$ then $b \in \U'$.

It is easy to check that $\U'$ is finite. Indeed, since
$(\U,\astore,\afunc) \models \alpha$ and no new node or edge is added,
all nodes are of height less or equal to $n+1$. Furthermore, all nodes
have at most $\card{B} +\card{\Pi} + \#K$ children in $\U'$, where
$\#K$ denotes the number of $E_n$-equivalence classes.

We denote by $\I' = (\U',\astore,\afunc')$ the restriction of $\I$ to
the elements of $\U'$ (we may assume that $\astore$ is a store on
$\U'$ since $\varphi$ is closed). We prove that $\I' \models \varphi$.
\begin{compactitem}
\item{ Since $\U'$ contains the root, and $\I \models\alpha$, we must
  have $\I' \models \alpha$.  
}
\item{ Observe that $\U'$ necessarily contains $\nodesCh(b,K)$, for
  every $b \in \U'$, since by definition the parent of the (unique)
  element of $\nodesCh(b,K)$ is $b$.  Thus at least one child of $b$
  is kept in each equivalence class.  Thus the relations $E_m$ on
  elements of $\U'$ are preserved in the transformation: for every
  $a,b \in \U'$, $a,b$ are $E_m$-equivalent in the structure $\I$ iff
  they are equivalent in the structure $\I'$.  Further, the height of
  the nodes cannot change. Therefore, for every $a,a' \in U'$:
\[(\U',\astore[x \mapsto a,y\mapsto a'],\afunc') \models \beta_n(x,y) 
\text{ iff } (\U,\astore[x \mapsto a,y\mapsto a'],\afunc) \models
\beta_n(x,y)\] By definition, for every $a \in B$ and $m \in \nat$,
$\semf^m(a) \in \{ a,\rootnode \}$, thus $B \subseteq \U'$.  Because
no new edges are added, we deduce: \[(\U',\astore[x \mapsto
  a],\afunc') \models \delta(x) \iff (\U,\astore[x \mapsto a],\afunc)
\models \delta(x) \iff a \in B\] Consequently, since $\I \models
\gamma$, we have $\I' \models \gamma$.  }
\item{ All elements in $\nodesPi(i,a,a')$ with $a,a'\in B$ occur in
  $\U'$ (because if $b \in \nodesPi(i,a,a')$ and $m \in \nat$ then
  $\semf^m(b) \in \{ \rootnode \} \cup B \cup \nodesPi(i,a,a')$),
  thus, for all $a,a' \in B$:
\[(\U',\astore[x  \mapsto a,y\mapsto a'], \afunc')  \models  \pi_i(x,y) 
\iff (\U,\astore[x \mapsto a,y\mapsto a'],\afunc) \models \pi_i(x,y)\]
Since all quantifications in $\eta'$ range over elements in $B$, we
deduce, by a straightforward induction on the formula, that $\I$ and
$\I'$ necessarily agree on the formula
$\eta'[D(x)/\delta(x),P_i(x,y)/\pi_i(x,y)]$. Consequently, we must
have $\I' \models \eta'[D(x)/\delta(x),P_i(x,y)/\pi_i(x,y)]$. \qed}
\end{compactitem} 
}

\begin{theorem}\label{thm:sl1nonelementary}
  The finite and infinite satisfiability problems are not elementary
  recursive for prenex formulae of $\seplogk{1}$.
\end{theorem}
\ifLongVersion
\proof{ The lower bound complexity result follows from the complexity
  result of Lemma \ref{lem:book} and from the reductions in
  Propositions \ref{prop:fin} and \ref{prop:infin}. \qed}
\fi

\section{The $\bsr(\seplogk{1})$ Fragment is \pspace-complete}

The last result concerns the tight complexity of the
$\bsr(\seplogk{1})$ fragment. For $k\geq2$, we showed that
$\bsr(\seplogk{k})$ is undecidable, in general, and \pspace-complete
if the positive occurrences of the magic wand are
forbidden\footnote{For infinite satisfiability, it is enough to forbid
  positive occurrences of the magic wand containing universally
  quantified variables only.}. Here we answer the problem concerning
the exact complexity of $\bsr(\seplogk{1})$, by showing
its \pspace-completeness. 

Let $\I = (\U,\astore,\aheap)$ be a structure, $X$ a non-empty set of
variables and $L \subseteq \dom(\aheap)$ be a set of locations. We
define:
\[\begin{array}{rcl}
V_{X,L} & \isdef & L \cup \astore(X) \\
\overline{V}_{X,L} & \isdef & \set{\ell \in \U \mid 
\exists i \geq 0 ~\exists \ell' \in V_{X,L} ~.~ \aheap^i(\ell') = \ell} \\
W_{X,L} & \isdef & V_{X,L} \cup \set{\ell \in \overline{V}_{X,L} \mid 
  \exists \ell',\ell'' \in \overline{V}_{X,L} ~.~ \ell' \neq \ell'' 
  \wedge \aheap(\ell') = \aheap(\ell'') = \ell}
\end{array}\]
Intuitively, $\overline{V}_{X,L}$ contains all locations reachable via
the heap from a location either in $L$ or labelled with a variable
from $X$ and $W_{X,L}$ contains all locations from $V_{X,L}$ and those
from $\overline{V}_{X,L}$ that have two or more predecessors via the
heap. 

Given a location $\ell_0 \in \dom(\aheap)$, the \emph{segment}
$S(\ell_0)=\tuple{\ell_0,\ell_1,\ldots,\ell_n}$, for some $n\geq0$, is
the unique sequence of locations such that $\ell_1,\ldots,\ell_n \in
\dom(\aheap) \setminus W_{X,L}$, $\aheap(\ell_i)=\ell_{i+1}$ for all
$i \in [0,n-1]$ and either $\aheap(\ell_n)\in W_{X,L}$ or
$\aheap^2(\ell_n)=\bot$. Note that because the domain of $\aheap$ is
necessarily finite, such a sequence is well defined. We denote by
$\len{S(\ell_0)}=n+1$ the number of locations in the segment. For an
integer $N\geq0$, we denote by $S^N(\ell_0)$ the restriction of
$S(\ell_0)$ to its first $\min(\len{S(\ell_0)}-1,N)+1$ elements. We
sometimes blur the distinction between a segment and the set of its
elements and write $\ell \in S(\ell_0)$ iff $\ell$ is one of the
elements of $S(\ell_0)$.

Given a structure $\I=(\U,\astore,\aheap)$, the
\emph{$(N,X,L)$-contraction of $\I$} is the structure $C^N_{X,L}(\I) =
(\U',\astore,\aheap')$ defined as follows: \begin{compactitem}
\item $\U' \isdef (\U \setminus \overline{V}_{X,L}) \cup
  \bigcup_{\ell_0 \in W_{X,L}} S^N(\ell_0)$,
\item for each $\ell \in (\U \setminus \overline{V}_{X,L}) \cup
  W_{X,L}$,  $\aheap'(\ell) \isdef \aheap(\ell)$,
\item for each $\ell_0 \in W_{X,L}$ such that $S^N(\ell_0) =
  \tuple{\ell_0,\ldots,\ell_M}$ and $M=\min(\len{S(\ell_0)}-1,N)$,
  we define: \begin{compactitem}
  \item $\aheap'(\ell_i) \isdef \aheap(\ell_i) ~[= \ell_{i+1}]$ for
    all $i \in [1,M-1]$, and
  \item $\aheap'(\ell_M) \isdef \aheap^i(\ell_M)$, where $i>0$ is the
    smallest integer such that either $\aheap^i(\ell_M) \in W_{X,L}$ or
    $\aheap^{i+1}(\ell_M) = \bot$. Such an integer necessarily exists by 
    definition of $S(\ell_0)$.
  \end{compactitem}
\end{compactitem}

\begin{proposition}\label{prop:contraction-size}
  Given a structure $\I=(\U,\astore,\aheap)$, for any
  $(N,X,L)$-contraction $C^N_{X,L}(\I)=(\U',\astore,\aheap')$, we have
  $\card{\U'} - \card{(\U\setminus\overline{V}_{X,L})} \leq
  2N(\card{\astore(X)}+\card{L})$.
\end{proposition}
\proof{ By induction on $\card{V_{X,L}}\geq1$, one shows that
  $\card{W_{X,L}\setminus V_{X,L}} \leq \card{V_{X,L}}$, which implies
  $\card{W_{X,L}\setminus V_{X,L}} \leq
  \card{\astore(X)}+\card{L}$. If $\card{V_{X,L}}=1$ then there exists
  at most one location $\ell \in W_{X,L}$ such that $\ell =
  \aheap^i(\ell_0) = \aheap^j(\ell)$, for some $\ell_0 \in V_{X,L}$
  and some $i,j > 0$. Thus $\card{W_{X,L}\setminus V_{X,L}}\leq1$. Let
  $\ell_0 \in V_{X,L}$ be a location, $V^0_{X,L} = V_{X,L} \setminus
  \set{\ell_0}$ and $\overline{V^0}_{X,L}$, $W^0_{X,L}$ be the sets
  defined using $V^0_{X,L}$ instead of $V_{X,L}$. We distinguish the
  following cases: \begin{compactitem}
    \item If all locations reachable from $\ell_0$ are outside
      $\overline{V^0}_{X,L}$, then there exists at most one location
      $\ell$ such that $\ell = \aheap^i(\ell_0) = \aheap^j(\ell)$, for
      some $i,j > 0$, thus either $W_{X,L} = W^0_{X,L}$ or $W_{X,L} =
      W^0_{X,L} \cup \set{\ell}$.
    \item Otherwise, there exists a location $\ell \in
      \overline{V^0}_{X,L}$ such that $\ell = \aheap^i(\ell_0)$, for
      some $i > 0$ and let $i$ be the minimal such number. Then we
      have $W_{X,L} = W^0_{X,L} \cup \set{\ell}$.
  \end{compactitem}
  In both cases we have $W_{X,L} \subseteq W^0_{X,L} \cup \set{\ell}$, for some location $\ell$. 
  We compute: \[\begin{array}{rcl}
  W_{X,L} \setminus V_{X,L} & \subseteq & W_{X,L} \setminus V^0_{X,L} \\
  & \subseteq & (W^0_{X,L} \cup \set{\ell}) \setminus V^0_{X,L} \\
  & = & (W^0_{X,L} \setminus V^0_{X,L}) \cup (\set{\ell} \setminus V^0_{X,L})
  \end{array}\]
  Then we obtain:
  \[\begin{array}{rcl}
  \card{W_{X,L} \setminus V_{X,L}} & \leq & \card{W^0_{X,L} \setminus V^0_{X,L}} + \card{\set{\ell} \setminus V^0_{X,L}} \\
  & \leq & \card{W^0_{X,L} \setminus V^0_{X,L}} + 1 \\
  \text{(induction hypothesis) } & \leq & \card{V^0_{X,L}} + 1 \leq \card{V_{X,L}}
  \end{array}\]  
  Since every segment in $C_{N,X,L}$ has length at most $N$, we obtain
  that $\U'$ contains at most $\card{(\U\setminus\overline{V}_{X,L})}
  + 2N(\card{\astore(X)}+\card{L})$ locations. \qed}

\begin{lemma}\label{lemma:contraction-small-model}
  Let $\psi = \exists y_1 \ldots \exists y_m ~.~ \phi(x_1,\ldots,x_n,
  y_1,\ldots,y_m)$ be a formula, where such that $n,m\geq1$ and $\phi$
  is a quantifier-free boolean combination of test formulae. Let $X =
  \set{x_1,\ldots, x_n}$ and consider a structure $\I =
  (\U,\astore,\aheap)$ such that there exists a set of locations $L
  \subseteq \U$ with $\card{L\cap\dom(\aheap)} \geq \maxn{\phi}$. If
  $C^{m}_{X,L}(\I) \models \psi$ then $\I \models \psi$.
\end{lemma}
\proof{ Let $C^{m}_{X,L}(\I) = (\U',\astore,\aheap')$. If
  $(\U',\astore,\aheap') \models \psi$ then there exists a sequence of
  locations $\ell'_1, \ldots, \ell'_m \in \U'$ such that
  $(\U',\astore[y_1 \leftarrow \ell'_1,\ldots,y_m \leftarrow
    \ell'_m],\aheap') \models \phi$. We shall build a sequence
  $\ell_1,\ldots,\ell_m \in \U$ such that $(\U,\astore[y_1 \leftarrow
    \ell_1,\ldots,y_m \leftarrow \ell_m],\aheap) \models
  \phi$. Initially, for each $\ell'_i \in (\U \setminus
  \overline{V}_{X,L}) \cup W_{X,L}$, let $\ell_i \isdef \ell'_i$ and
  mark the index $i$ as visited. Then repeat the following steps,
  until there are no more unmarked indices in
  $[1,m]$: \begin{compactenum}
  \item\label{it1:contraction} For each unmarked index $i$ such that
    $\ell'_i=\ell'_j$ for some marked index $j$, let $\ell_i \isdef
    \ell_j$ and mark $i$.
  \item\label{it2:contraction} Choose an unmarked index $i$. Since $i$
    is unmarked, necessarily $\ell'_i \not\in (\U \setminus
    \overline{V}_{X,L}) \cup W_{X,L}$ hence $\ell'_i\in
    S^{m}(\ell'_0)$, for some $\ell'_0 \in W_{X,L}$. Let ${i_1} <
    \ldots < {i_q}$ be the set of unmarked indices such that
    $\ell'_{i_1}, \ldots, \ell'_{i_q} \in S^{m}(\ell'_0)$, and
    consider the numbers $r_1,\ldots, r_q$ such that:
    \begin{equation}\label{eq:contraction}
      {\aheap'}^{r_1}(\ell'_0) = \ell'_{i_1}, \ldots, 
      {\aheap'}^{r_{j+1}}(\ell'_{i_j}) = \ell'_{i_{j+1}}, \ldots,
      {\aheap'}^{r_q}(\ell'_{i_q}) = \aheap^t(\ell'_0)
    \end{equation} 
    where $t>0$ is the smallest number such that either
    $\aheap^t(\ell'_0) \in W_{X,L}$ or $\aheap^{t+1}(\ell'_0) =
    \bot$. Note that in particular $\sum_{i=1}^q r_i = t$. If $t \leq
    m$ then let $\ell_{i_j} \isdef \ell'_{i_j}$ for all $j \in
    [1,q]$. Otherwise, since $r_1+\ldots+r_q > m$ and $q \leq m$,
    there exists $h\in[1,q]$ such that $r_h \geq 2$. Let $h$ be the
    maximal such number. Then let $\ell_{i_j} \isdef \ell'_{i_j}$ if
    $j \in [1,h]$ and $\ell_{i_j} \isdef \aheap^{t-\sum_{s=j}^q
      r_s}(\ell'_0)$ if $j \in [h+1,q]$. Finally mark ${i_1}, \ldots,
    {i_q}$ as visited.
  \end{compactenum}
  Now we show that, for any literal $\lambda$, if
  $(\U',\astore[y_1 \leftarrow \ell'_1,\ldots,y_m \leftarrow
    \ell'_m],\aheap') \models \lambda$ then $(\U,\astore[y_1
    \leftarrow \ell_1,\ldots,y_m \leftarrow \ell_m],\aheap) \models
  \lambda$, by a case split on the form of
  $\lambda$: \begin{compactitem}
  \item $x \teq y$, $\neg x \teq y$ 
    \begin{compactitem} 
    \item If $x,y \in X$ then $\astore[y_1 \leftarrow
      \ell'_1,\ldots,y_m \leftarrow \ell'_m]$ and $\astore[y_1
      \leftarrow \ell_1,\ldots,y_m \leftarrow \ell_m]$ agree on the
      values assigned to $x$ and $y$.
    \item If $x \in X$ and $y = y_i$ for some $i \in [1,m]$ then
      $\astore(x) \in V_{X,L}$. If we also have $\ell'_i \in V_{X,L}$
      then $\ell_i \isdef \ell'_i$ and $\astore[y_1 \leftarrow
        \ell'_1,\ldots,y_m \leftarrow \ell'_m]$ and $\astore[y_1
        \leftarrow \ell_1,\ldots,y_m \leftarrow \ell_m]$ agree on the
      values assigned to $x$ and $y$ because both values are in
      $W_{X,L}$. Otherwise, $\ell'_i \not\in V_{X,L}$ and suppose, by
      contradiction, that $\ell_i \in V_{X,L}$. We distinguish the
      following cases: \begin{compactitem}
      \item if $\ell_i$ is assigned initially, then we have $\ell_i =
        \ell'_i \not\in V_{X,L}$, contradiction. 
      \item else, if $\ell_i$ is assigned at step
        \ref{it2:contraction}, it is necessarily assigned to some
        location not in $V_{X,L}$, contradiction.
      \item otherwise, if $\ell_i$ is assigned to some $\ell_j$ (step
        \ref{it1:contraction}) because $\ell'_i = \ell'_j$ then we
        obtain $\ell_j \in V_{X,L}$, $\ell'_j \not\in V_{X,L}$ and the
        argument is repeated inductively, until a contradiction is
        reached.
      \end{compactitem}
      Then the values assigned to $x$ and $y$ are different for both
      $\astore[y_1 \leftarrow \ell'_1,\ldots,y_m \leftarrow \ell'_m]$
      and $\astore[y_1 \leftarrow \ell_1,\ldots,y_m \leftarrow
        \ell_m]$.
    \item Otherwise, $x=y_i$ and $y=y_j$ for some $i,j \in [1,m]$. Then
      $\ell_i=\ell_j$ iff $\ell'_i=\ell'_j$, by definition (step
      \ref{it1:contraction}).
  \end{compactitem}
  \item $\alloc(x)$: \begin{compactitem}
    \item If $x \in X$, then since $\astore(x) \in \dom(\aheap')$ we must
      have $\astore(x) \in \dom(\aheap)$, because $\dom(\aheap')
      \subseteq \dom(\aheap)$.
    \item Otherwise $x = y_i$ for some $i \in [1,m]$ and $\ell'_i \in
      \dom(\aheap')$. We distinguish the following cases, based on the
      definition of $\ell_i$: \begin{compactitem}
      \item if $\ell_i$ is assigned initially, we have $\ell_i = \ell'_i \in \dom(\aheap') \subseteq \dom(\aheap)$, 
      \item else, if $\ell_i$ is assigned at step
        \ref{it2:contraction} then necessarily $\ell_i \in
        \dom(\aheap)$,
      \item otherwise, if $\ell_i$ is assigned to some $\ell_j$ (step
        \ref{it1:contraction}) because $\ell'_i = \ell'_j$ then we are
        left with proving $\ell_j \in \dom(\aheap)$, repeating the
        argument inductively.
      \end{compactitem}
    \end{compactitem}
  \item $\neg\alloc(x)$: \begin{compactitem}
    \item If $x \in X$ then $\astore(x) \in V_{X,L} \subseteq
      W_{X,L}$. By construction, $\dom(\aheap') \cap W_{X,L} =
      \dom(\aheap) \cap W_{X,L}$, thus $\astore(x) \not\in
      \dom(\aheap')$ implies $\astore(x) \not\in \dom(\aheap)$.
    \item Otherwise $x = y_i$ for some $i \in [1,m]$ and $\ell'_i \not\in
      \dom(\aheap')$. Then either $\ell'_i \in (\U \setminus
      \overline{V}_{X,L}) \cup W_{X,L}$, in which case $\ell_i =
      \ell'_i$ by definition and $\dom(\aheap') \cap [(\U \setminus
        \overline{V}_{X,L}) \cup W_{X,L}] = \dom(\aheap) \cap [(\U
        \setminus \overline{V}_{X,L}) \cup W_{X,L}]$, or $\ell'_i \in
      S^m(\ell_0)$ for some $\ell_0 \in W_{X,L}$. The latter case,
      however, contradicts the fact that $\ell'_i \not\in \dom(\aheap')$.
    \end{compactitem}
  \item $x \pto y$: \begin{compactitem}
  \item If $x,y \in X$, then since $\aheap'(\astore(x))=\astore(y)$ and
    $\astore(x) \in W_{X,L}$, we have $\aheap(\astore(x))=\astore(y)$
    because $\aheap'$ agrees with $\aheap$ on $W_{X,L}$.
  \item If $x \in X$ and $y=y_i$ for some $i \in [1,m]$, we have
    $\aheap'(\astore(x))=\aheap(\astore(x))=\ell'_i$, because
    $\astore(x) \in W_{X,L}$ and $\aheap'$ agrees with $\aheap$ on
    $W_{X,L}$. There remains to show that $\ell'_i=\ell_i$ in this
    case. If $\ell'_i \in (\U \setminus \overline{V}_{X,L}) \cup
    W_{X,L}$ then this is the case by definition. Otherwise $\ell'_i
    \in S^m(\astore(x))$. Thus, $r_1 = 1$, where $r_1,\ldots,r_q$ is
    the sequence of numbers in step \ref{it2:contraction} of the
    construction above. If $t\leq m$, then $l_i = l_i'$ by
    construction. Otherwise, since $r_1 = 1$, the maximal number $h$
    such that $r_h\geq 2$ is strictly greater than $1$ and once again,
    $l_i = l_i'$.
%Thus $r_1 = 1$ and $\ell_i = \ell'_i$ by
%    definition, where $r_1,\ldots,r_q$ (\ref{eq:contraction}) is the
%    sequence of numbers from the definition of
%    $\ell'_1,\ldots,\ell'_m$ (step \ref{it2:contraction}).
    %
  \item If $x=y_i$ for some $i \in [1,m]$ and $y\in X$, we have
    $\aheap'(\ell'_i)=\astore(y) \in W_{X,L}$. We distinguish the
    following cases: \begin{compactitem}
    \item If $\ell'_i \in (\U \setminus \overline{V}_{X,L}) \cup
      W_{X,L}$ then $\ell_i=\ell'_i$ by definition and moreover
      $\aheap'$ agrees with $\aheap$ on $(\U \setminus
      \overline{V}_{X,L}) \cup W_{X,L}$. 
    \item Otherwise $\ell'_i \in S^m(\ell'_0)$ for some $\ell'_0 \in
      W_{X,L}$. Since $\astore(y) \in W_{X,L}$ it must be that
      $\ell'_i$ is the last location in $S^m(\ell'_0)$, hence $r_q=1$,
      where $r_1,\ldots,r_q$ (\ref{eq:contraction}) is the sequence of
      numbers from the definition of $\ell'_1,\ldots,\ell'_m$ (step
      \ref{it2:contraction}). Then either $r_j = 1$ for all $j \in
          [1,q]$, in which case $\ell'_i = \ell_i$, or $r_h \geq 2$
          for some $h \in [1,q]$. However, since $h \neq q$, we also
          have that $\ell'_i = \ell_i$ in this case.
    \end{compactitem}
  \item If $x=y_i$ and $y=y_j$, for some $i,j\in[1,m]$, we have
    $\aheap'(\ell'_i)=\ell'_j$ and we prove that
    $\aheap(\ell_i)=\ell_j$ as well. We distinguish the following
    cases: \begin{compactitem}
    \item If $\ell'_i,\ell'_j \in (\U \setminus \overline{V}_{X,L})
      \cup W_{X,L}$ then since $\ell'_i=\ell_i$, $\ell'_j=\ell_j$
      and $\aheap'$, $\aheap$ agree on $W_{X,L}$, we have the result.
  \item Otherwise, if $\ell'_i \in S^m(\ell_0)$, for some $\ell_0 \in
    W_{X,L}$, let $r_p=1$ be the number such that
    ${\aheap'}^{r_p}(\ell'_i)=\ell'_j$ in (\ref{eq:contraction}),
    where $r_1,\ldots,r_q$ (\ref{eq:contraction}) is the sequence of
    numbers from the definition of $\ell'_1,\ldots,\ell'_m$ (step
    \ref{it2:contraction}). Then either $r_j = 1$ for all $j \in
        [1,q]$, in which case $\ell'_i = \ell_i$ and $\ell'_j =
        \ell_j$, or $r_h \geq 2$ for some $h \in [1,q]$. However,
        since $h \neq p$, we also have that $\ell'_i = \ell_i$ and
        $\ell'_j = \ell_j$, in this case.
    \end{compactitem}
  \end{compactitem}
  \item $\neg x \pto y$: If $\astore(x) \not\in \dom(\aheap')$ we show
    that $\astore(x) \not\in \dom(\aheap)$, as in the $\neg \alloc(x)$
    case above. Otherwise, $\astore(x) \in \dom(\aheap')$ and
    $\aheap'(\astore(x)) \neq \astore(y)$. We distinguish the
    following cases: \begin{compactitem}
    \item $x,y \in X$ is similar to the case $x \pto y$ for $x,y \in
      X$, above.
    \item If $x \in X$ and $y = y_i$, for some $i \in [1,m]$, we have
      $\aheap'(\astore(x)) = \aheap(\astore(x)) \neq \ell'_i$, because
      $\astore(x) \in W_{X,L}$ and $\aheap'$, $\aheap$ agree on
      $W_{X,L}$. Suppose, by contradiction, that
      $\aheap(\astore(x))=\ell_i$. Then $\ell_i \in S(\astore(x)) =
      \langle \astore(x), \ell_i, \ldots \rangle$ and since $m \geq
      1$, also $\ell_i \in S^m(\astore(x))$, which leads to
      $\ell_i=\ell'_i$, in contradiction with $\aheap'(\astore(x))
      \neq \ell'_i$.
    \item If $x=y_i$ for some $i \in [1,m]$ and $y \in X$, then
      $\aheap'(\ell'_i)\neq\astore(y)$ and suppose, by contradiction,
      that $\aheap(\ell_i)=\astore(y)$. We distinguish the following
      cases: \begin{compactitem}
      \item If $\ell_i \in (\U \setminus \overline{V}_{X,L}) \cup
        W_{X,L}$ then $\ell_i=\ell'_i$ by definition and moreover
        $\aheap'$ agrees with $\aheap$ on $\ell'_i \in (\U \setminus
        \overline{V}_{X,L}) \cup W_{X,L}$, which contradicts with
        $\aheap'(\ell'_i)\neq\astore(y)$.
      \item Otherwise $\ell_i \in S(\ell_0)$ for some $\ell_0 \in
        W_{X,L}$ and since $\astore(y) \in W_{X,L}$, we have $r_q=1$
        and $\ell'_i = \ell_i$, by definition, where $r_1,\ldots,r_q$
        (\ref{eq:contraction}) is the sequence of numbers from the
        definition of $\ell'_1,\ldots,\ell'_m$ (step
        \ref{it2:contraction}). Moreover, $\aheap'(\ell'_i) =
        \astore(y)$ by the definition of $\aheap'$, which contradicts
        with $\aheap'(\ell'_i)\neq\astore(y)$.
      \end{compactitem}
    \item If $x = y_i$ and $y = y_j$, for some $i,j \in [1,m]$, such
      that $\aheap'(\ell'_i) \neq \ell'_j$. Suppose, by contradiction,
      that $\aheap(\ell_i)=\ell_j$. We distinguish the following
      cases: \begin{compactitem}
      \item if $\ell_i,\ell_j \in (\U \setminus \overline{V}_{X,L})
        \cup W_{X,L}$ then $\ell'_i=\ell_i$, $\ell'_j=\ell_j$ and
        $\aheap'$ and $\aheap$ agree on $(\U \setminus
        \overline{V}_{X,L}) \cup W_{X,L}$, then $\aheap'(\ell'_i) =
        \ell'_j$, contradiction.
      \item if $\ell_i \in (\U \setminus \overline{V}_{X,L}) \cup
        W_{X,L}$ and $\ell_j \not\in (\U \setminus \overline{V}_{X,L})
        \cup W_{X,L}$, then $\ell'_i=\ell_i$,
        $\aheap'(\ell'_i)=\aheap(\ell_i)$ and $S(\ell'_i) = \langle
        \ell'_i, \ell_j, \ldots \rangle$. But then $r_1=1$
        (\ref{eq:contraction}) and $\ell'_j=\ell_j$ by definition,
        contradiction.
      \item if $\ell_i \not\in (\U \setminus \overline{V}_{X,L}) \cup W_{X,L}$ and
        $\ell_j \in (\U \setminus \overline{V}_{X,L}) \cup W_{X,L}$, then
        $\ell'_j=\ell_j$ and $\ell_i \in S(\ell_0)$ for some $\ell_0
        \in W$. But then $\ell_i$ is the last location in the segment,
        thus $r_q=1$ (\ref{eq:contraction}) and $\ell'_i=\ell_i$,
        $\aheap'(\ell'_i)=\ell'_j$ follows, contradiction. 
      \item if $\ell_i \not\in (\U \setminus \overline{V}_{X,L}) \cup
        W_{X,L}$ and $\ell_j \not\in (\U \setminus \overline{V}_{X,L})
        \cup W_{X,L}$, then $\ell_i,\ell_j \in S(\ell_0)$ for some
        $\ell_0 \in W_{X,L}$ and, moreover, $\ell_i$ and $\ell_j$ are
        consequtive locations in $S(\ell_0)$. Then the only
        possibility is that $\ell'_i,\ell'_j \in S^m(\ell_0)$ and
        $\aheap'(\ell'_i)=\ell'_j$, contradiction.
      \end{compactitem}
    \end{compactitem}
  \item $\len{h} \geq \len{U}-n$, $\len{h} < \len{U}-n$: Let $T =
    \bigcup_{\ell_0\in W_{X,L}} S(\ell_0) \setminus S^m(\ell_0)$. It
    is not hard to show that \begin{inparaenum}[(i)]
    \item $T \subseteq \dom(\aheap)$,
    \item $\U' = \U \setminus T$ and
    \item $\dom(\aheap') = \dom(\aheap) \setminus T$.
    \end{inparaenum}    
    Then $\card{\U}-\card{\dom(\aheap)} = \card{\U'} -
    \card{\dom(\aheap')}$ and the result follows.
  \item $\len{h}\geq n$: we have $\card{\dom(\aheap)} \geq
    \card{\dom(\aheap')} \geq n$.
  \item $\len{h} < n$: since $\card{L\cap\dom(\aheap)} \geq
    \maxn{\varphi}$, we have $\card{\dom(\aheap')} \geq n$, thus $\I'
    \not\models \len{h} < n$.
  \item $\len{U}\geq n$: we have $\card{\U} \geq \card{\U'} \geq n$. 
  \item $\len{U} < n$: since $\card{L\cap\dom(\aheap)} \geq
    \maxn{\varphi}$, we have $\card{\U'} \geq n$, thus $\I'
    \not\models \len{U} < n$.
  \end{compactitem}
\qed}

Given a set $L \subseteq \U$, the \emph{$(X,L)$-restriction}
$R_{X,L}(\I) = (\U',\astore,\aheap')$ is defined as $\U' \isdef
\overline{V}_{X,L}$, and for each $\ell \in \U'$, $\aheap'(\ell)
\isdef \aheap(\ell)$. Observe that, because $\overline{V}_{X,L}$ is
closed under applications of $\aheap$, we have $\dom(\aheap') \cup
\img(\aheap') \subseteq \U'$. 

%% \begin{proposition}\label{prop:restriction-sound}
%%   Given a structure $\I = (\U,\astore,\aheap)$, for any
%%   $(N,L)$-restriction $R_{X,L}(\I) = (\U',\astore,\aheap')$ we have
%%   $\elemh{\aheap'} \subseteq \U'$.
%% \end{proposition}
%% \proof{ Clearly $\dom(\aheap') = \U' \cap \dom(\aheap)$. Suppose, for
%%   a contradiction, that $\img(\aheap') \not\in \U'$, i.e.\ there
%%   exists $\ell \in \dom(\aheap')$ such that $\aheap'(\ell) \not\in
%%   \U'$. Then $\aheap'(\ell) = \aheap(\ell) \in
%%   \widetilde{V}_{X,L}$. Since $\ell \in \dom(\aheap)$, it is enough to
%%   show that $\ell \in \widetilde{V}_{X,L}$, by proving the
%%   contrapositive $\ell \in \overline{V}_{X,L} \Rightarrow \aheap(\ell)
%%   \in \overline{V}_{X,L}$. If $\ell \in \overline{V}_{X,L}$ there
%%   exists $\ell'\in V_{X,L}$ such that $\aheap^i(\ell')=\ell$, for some
%%   $i\geq0$. But then $\aheap(\ell)=\aheap^{i+1}(\ell') \in
%%   \overline{V}_{X,L}$. \qed}

\begin{lemma}\label{lemma:restriction-small-model}
  Let $\psi = \exists y_1 \ldots \exists y_m ~.~ \phi(x_1,\ldots,x_n,
  y_1,\ldots,y_m)$ be a formula, where $\phi$ is a quantifier-free
  boolean combination of test formulae with free variables
  $x_1,\dots,x_n,y_1,\dots,y_m$. Let $X = \set{x_1,\ldots, x_n}$ and
  consider a structure $\I = (\U,\astore,\aheap)$ such that there
  exists a set of locations $L \subseteq \U$ with $\card{L \cap
    \dom(\aheap)} \geq \maxn{\phi}$ and \(\card{(L \cup \astore(X))
    \setminus \dom(\aheap)} = \min(\card{U\setminus
    \dom(\aheap)},\maxn{\phi}+1)\). If $R_{X,L}(\I) \models \psi$ then
  $\I \models \psi$.
\end{lemma}
\proof{ If $R_{X,L}(\I) \models \psi$ then there exist
  $\ell_1,\ldots,\ell_m \in \U'$ such that
  $(\U',\astore[y_1\leftarrow\ell_1,\ldots,y_m\leftarrow\ell_m],\aheap')
  \models \phi$. We show that, for each literal $\lambda$, we have
  $(\U',\astore[y_1\leftarrow\ell_1,\ldots,y_m\leftarrow\ell_m],\aheap')
  \models \lambda \Rightarrow
  (\U,\astore[y_1\leftarrow\ell_1,\ldots,y_m\leftarrow\ell_m],\aheap)
  \models \lambda$, using a case split on the form of
  $\lambda$: \begin{compactenum}
  \item $x \teq y$, $\neg x \teq y$: trivial, because the store does
    not change between $\I'$ and $\I$.
  \item $\alloc(x)$: $\astore(x) \in \dom(\aheap') \subseteq
    \dom(\aheap)$.
  \item $\neg\alloc(x)$: $\astore(x) \in \U'\setminus\dom(\aheap')$
    \comment[me]{What do we do with the variables that are not
      allocated? Their image still needs to be in
      $\U'$}\comment[ri]{since $\U' = \overline{V}_{X,L}$ we have
      $\astore(x) \in \U'$ for all $x$ no ?}  and suppose that
    $\astore(x) \in \dom(\aheap)$. Since $\dom(\aheap') = \dom(\aheap)
    \cap \U'$, it must be the case that $\astore(x) \not\in \U'$,
    contradiction.
  \item $x \pto y$: $\astore(x), \astore(y) \in \U'$, $\astore(x) \in
    \dom(\aheap')$ and $\aheap'$ agrees with $\aheap$ on $\U'$.
  \item $\neg x \pto y$: if $\astore(x) \in \dom(\aheap')$ then
    $\astore(x) \in \dom(\aheap)$ and $\aheap(\astore(x)) =
    \aheap'(\astore(x))$, otherwise $\astore(x) \not\in \dom(\aheap')$
    and $\astore(x) \not\in \dom(\aheap)$ follows, by the argument
    used in the $\neg\alloc(x)$ case.
  \item $\len{h} \geq \len{U} - n$: $\card{\U' \setminus
    \dom(\aheap')} \leq n$ and, since $\U'=\overline{V}_{X,L}$ and
    $\dom(\aheap') = \dom(\aheap) \cap \overline{V}_{X,L}$, we
    compute:
    \[\begin{array}{rcl}
    \U' \setminus \dom(\aheap') & = & \overline{V}_{X,L} \setminus (\dom(\aheap) \cap \overline{V}_{X,L}) \\
    & = & \overline{V}_{X,L} \setminus \dom(\aheap) \\
    & \supseteq & (L \cup  \astore(X)) \setminus \dom(\aheap)
    \end{array}\]
    thus $\card{(L \cup \astore(X)) \setminus \dom(h)} \leq \card{\U'
      \setminus \dom(\aheap')} \leq n$, hence, since $n <
    \maxn{\phi}+1$, we have $\card{(L \cup \astore(X)) \setminus
      \dom(h)} = \card{\U \setminus \dom(\aheap)} \leq n$.
  \item $\len{h} < \len{U} - n$: we have $\card{\U' \setminus
    \dom(\aheap')} > n$. Since $\U' \subseteq \U$ and $\dom(\aheap') =
    \dom(\aheap) \cap \U'$ this entails that $\card{\U \setminus
      \dom(\aheap)} > n$.     

%  \item $\len{h} \geq \len{U} - n$: since $\card{\U'} -
%    \card{\dom(\aheap')} = \card{\U' - \dom(\aheap')} \geq \card{L
%      \setminus \dom(\aheap')} \geq \card{L \setminus \dom(\aheap)} =
%    \maxn{\phi}+1 > n$, we cannot have $\card{\dom(\aheap')} \geq
%    \card{\U'} - n$.
    %
%  \item $\len{h}  < \len{U} -  n$: $\card{\U} -  \card{\dom(\aheap)} =
%    \card{\U   \setminus    \dom(\aheap)}   \geq   \card{L   \setminus
%      \dom(\aheap)} = \maxn{\phi}+1  > n$, thus $\card{\dom(\aheap)} <
%    \card{\U} - n$ holds.
    %
  \item $\len{h}\geq n$, $\len{h} < n$, $\len{U}\geq n$ and
    $\len{U}<n$: using the same argument as in the proof of Lemma
    \ref{lemma:contraction-small-model}. \qed
  \end{compactenum}
}

\begin{theorem}
  The finite and infinite satisfiability problems for
  $\bsr(\seplogk{1})$ are \pspace-complete.
\end{theorem}
\proof{
  \pspace-hardness follows from the proof that satisfiability of the
  quantifier free fragment of $\seplogk{2}$ is \pspace-complete
  \cite[Proposition 5]{CalcagnoYangOHearn01}. This proof does not
  depend on the universe being infinite or $k=2$. It remains to
  show \pspace-membership for both problems.

  Let $\psi = \forall y_1 \ldots \forall y_m ~.~
  \phi(x_1,\ldots,x_n,y_1,\ldots,y_m)$, where $\phi$ is a
  quantifier-free $\seplogk{1}$ formula with free variables
  $x_1,\dots,x_n,y_1,\dots,y_m$.  By Lemma \ref{lem:infinitetofinite},
  $\psi$ has an infinite model iff $\psi \wedge \form{n+m}$ has a
  finite model, where the size of $\form{n+m}$ is quadratic
  in $n+m$.  Moreover, since $\form{n+m}$ is a $\bsr(\seplog)$ formula, $\psi \wedge \form{n+m}$ is a $\bsr(\seplog)$ formula.  We may
  therefore focus on the finite satisfiability problem.
  
  By Proposition \ref{prop:equivs}, $\psi$ has a finite model iff it
  has a model $\I = (\U,\astore,\aheap)$ such that $\card{\U \setminus
    \elemh{\aheap}} \leq m + n$.  Suppose that $\I \models \psi$ where
  $\I=(\U,\astore,\aheap)$ and $\card{\U \setminus \elemh{\aheap}}
  \leq m + n$. We prove that $\psi$ has a model % \wedge \form{m}$ 
  $\I'=(\U',\astore,\aheap')$ such that $\card{\aheap'} \leq
  \card{\U'} \leq \mathcal{P}(\len{\varphi})$, for some polynomial
  function $\mathcal{P}(x)$.
  
  Let $\mu = \bigvee_{M \in \finmt{\phi}} M$ be the expansion of
  $\phi$ as a disjunction of minterms that preserves all its finite
  models. By Lemma \ref{lemma:sl-mnf}, the formula $\psi$ is
  equivalent on finite models to $\forall y_1,\dots,y_m~.~ \mu$.  Let
  $X = \set{x_1,\ldots,x_n}$ and $N=\max_{M \in \finmt{\phi}}
  \maxn{M}$. If there is no set $L \subseteq \U \setminus \astore(X)$
  such that $\card{L \cap \dom(\aheap)} = N$ and $\card{(L \cup
    \astore(X)) \setminus \dom(\aheap)} = \min(\card{U\setminus
    \dom(\aheap)},N+1)$, then $\card{\dom(\aheap)} < N+n$ must be the
  case, as we show next. Suppose, by contradiction, that
  $\card{\dom(\aheap)} \geq N+n$. Then there exists a set $L_1
  \subseteq (\U \setminus \astore(X)) \cap \dom(\aheap)$ such that
  $\card{L_1} = N$.  Let $n' = \card{\astore(X) \setminus
    \dom(\aheap)}$.  By definition, $(\U \setminus \astore(X))
  \setminus \dom(\aheap)$ contains $\card{\U \setminus
    \dom(\aheap)}-n'$ elements.  Hence there exists a set $L_2
  \subseteq (\U \setminus \astore(X)) \setminus \dom(\aheap)$ such
  that $\card{L_2} = \min(\card{U\setminus
    \dom(\aheap)},\maxn{\phi}+1)-n'$.  Let $L\isdef L_1\cup L_2$. The
  sets $L_1$, $L_2$ and $\astore(X)$ are pairwise disjoint, and since
  $L_1\subseteq \dom(\aheap)$, we have $(L\cup \astore(X))\setminus
  \dom(\aheap) = L_2\cup (\astore(X)\setminus \dom(\aheap))$. We
  deduce that $\card{(L \cup \astore(X)) \setminus \dom(\aheap)} =
  \card{L_2}+n' = \min(\card{U\setminus \dom(\aheap)},\maxn{\phi}+1)$
  and $\card{L \cap \dom(\aheap)}= \card{L_1}=N$.
%By letting $L = L_1 \cup L_2$, 
%  we deduce that $\card{(L \cup \astore(X)) \setminus \dom(\aheap)} =
%  \card{L_2}+n' = \min(\card{U\setminus \dom(\aheap)},\maxn{\phi}+1)$
%  and $\card{L \cap \dom(\aheap)}= \card{L_1}=N$. 
  %Since $\card{X} = n$ and $\card{\dom(\aheap)} \geq N +
  %\card{\astore(X)}$, such a set must exist. Since $\card{L \cap
  %    \dom(\aheap)} = \card{L} = N$, the only possibility is $\card{L
  %    \setminus \dom(\aheap)} > \min(\card{U\setminus
  %    \dom(\aheap)},N+1)$. Since $L \setminus \dom(\aheap) = \emptyset$,
  %  by the choice of $L$, we end up with a contradiction.

  Hence $\card{\dom(\aheap)} < N+n$ and $\card{\elemh{\aheap}} <
  2(N+n)$, since each allocated location points to exactly one
  location, allocated or not. Therefore, $\card{\U} < m+n+2(N+n) =
  2N+3n+m$ and since $N$ is polynomially bounded by $\size{\varphi}$,
  by \cite[Lemma 7]{EIP18}, we are done, since we may assume that
  $\mathcal{P}$ is such that $\mathcal{P}(\len{\varphi}) \geq
  2N+3n+m$.

  Otherwise, let $L$ be such a set. By definition $\card{L} \leq
  N+N+1$. By Lemma \ref{lemma:restriction-small-model}, since $\I
  \models \forall y_1 \ldots \forall y_m ~.~ \mu$, we have
  $R_{X,L}(\I) \models \forall y_1 \ldots \forall y_m ~.~ \mu$ and by
  Lemma \ref{lemma:contraction-small-model}, we obtain
  $C^m_{X,L}(R_{X,L}(\I)) \models \forall y_1 \ldots \forall y_m ~.~
  \mu$.  Let $\I' = C^m_{X,L}(R_{X,L}(\I)) = (\U',\astore,\aheap')$
  and $\I'' = R_{X,L}(\I) = (\U'',\astore,\aheap'')$.  By definition
  of $R_{X,L}(\I)$, $\U'' = \overline{V}_{X,L}$.  By Proposition
  \ref{prop:contraction-size}, we have $\card{\U'} - \card{\U''
    \setminus \overline{V}_{X,L}} \leq 2m(n+\card{L})$, hence we
  deduce that $\card{\U'} \leq 2m(n+2N+1)$.  Again, the proof is
  completed, taking $\mathcal{P}(\len{\varphi}) = 2m(n+2N+1)$.

  We are left with proving that the model checking problem $\I \models
  \forall y_1 \ldots \forall y_m ~.~ \mu$ is in \pspace. We prove that
  the complement problem $\I \not\models \forall y_1 \ldots \forall
  y_m ~.~ \mu$ is in $\pspace$ and use the fact that $\pspace$ is
  closed under complement \cite[Corollary 4.21]{AroraBarak09}. Let
  $\I=(\U,\astore,\aheap)$. To check that $\I \models \exists y_1
  \ldots \exists y_m ~.~ \neg\mu$, we guess locations
  $\ell_1,\ldots,\ell_m \in \U$ and a \bounded~minterm $M$. Then we
  check that $M \in \finmt{\neg\psi}$ and that $(\U,\astore[y_1
    \leftarrow \ell_1, \ldots, y_m \leftarrow \ell_m], \aheap) \models
  M$. The first check is in \pspace, according to Lemma
  \ref{lemma:pspacemt} and the second is in \ptime. \qed}

\section{Conclusion}

We show that the prenex fragment of Separation Logic over heaps with
one selector, denoted as $\seplogk{1}$, is decidable in time not
elementary recursive. Moreover, the Bernays-Sch\"onfinkel-Ramsey
$\bsr(\seplogk{1})$ is \pspace-complete. These results answer an open
question raised in \cite{EIP18}, which established the undecidability
of $\seplogk{k}$, over heaps with $k\geq2$ selector fields.